\documentclass[aps,pre,twocolumn,superscriptaddress,showkeys,floatfix]{revtex4-2}
\usepackage{graphicx}
\usepackage{amsmath}
\usepackage{amsthm}
\usepackage{amssymb}
\usepackage{mathrsfs}

\newcommand{\mean}[1]{\left\langle #1 \right\rangle}   
\newcommand{\var}[1]{\mathrm{var}\!\left( #1 \right)}  
\newcommand{\dd}{\,\mathrm{d}}                         
\DeclareMathOperator{\ekurt}{\mathcal G}               
\DeclareMathOperator{\erg}{\mathcal U}                 
\DeclareMathOperator{\heat}{\mathcal C}                
\DeclareMathOperator{\free}{\mathcal F}                 
\DeclareMathOperator{\fK}{K}                           
\DeclareMathOperator{\fS}{S}                           

\begin{document}

\title{Revising the universality class of the four-dimensional Ising model}

\date{\today} 

\author{P. H. Lundow} 
\email{per.hakan.lundow@math.umu.se} 

\author{K. Markstr\"om}
\email{klas.markstrom@math.umu.se} 

\affiliation{Department of mathematics and mathematical statistics,
  Ume\aa{} University, SE-901 87 Ume\aa, Sweden}

\begin{abstract}
  The aim of this paper is to determine the behaviour of the specific
  heat of the 4-dimensional Ising model in a region aroud the critical temperature,
  and via that determine if the Ising model and the $\phi^4$-model
  belong to the same universality class in dimension 4.  In order to
  do this we have carried out what is currently the largest scale
  simulations of the 4-dimensional Ising model, extending the lattices
  size up to $L=256$ and the number of samples per size by several
  orders of magnitude compared to earlier works, keeping track of data
  for both the canonical and microcanonical ensembles.
	
  Our conclusion is that the Ising model has a bounded specific heat,
  while the $\phi^4$-model is known to have a logarithmic divergence
  at the critical point.  Hence the two models belong to distinct
  universality classes in dimension 4.
\end{abstract} 


\maketitle

\section{Introduction}
The 4-dimensional Ising model is of special import for, at least, two
reasons.  First, it is at the boundary, known as the upper critical
dimension, between the high-dimensional Ising models which follow
mean-field behaviour and the low-dimensional cases $D=1,2,3$ each of
which is qualitatively different from the others in just about every
interesting property.  Second, following the methods of constructive
field theory a well-behaved, i.e.\~ satisfying certain axioms,
4-dimensional spin model corresponds to one or several, depending on a
limit-taking procedure, time-dependent 3-dimensional quantum field
theories.  Hence a full understanding of the 4-dimensional Ising model
is desirable both in order to complete our understanding of
high-dimensional Ising models and as a necessary part of understanding
time-dependent quantum-field theory in 3-dimensional space.

There are still many basic questions regarding this model which remain
open. The broadest of these is arguably whether or not the Ising model
for $D=4$ belongs to the same universality class as the $\phi^4$-model
and, as a necessary condition for that inclusion, whether or not the
specific heat of the Ising model has the same type of singularity as
in the $\phi^4$-model.

Let us briefly recall that the lattice $\phi^4$-model is a spin model
similar to the Ising model but instead of having spins $\pm 1$, as in
the Ising model, the $\phi^4$-model allows all real numbers as spin
values and the action, which we can think as the equivalent of the
Hamiltonian in the Ising model, is $\beta
\sum_{ij}(\phi_i-\phi_j)^2+\sum_i(b \phi_i^2+\lambda \phi_i^4)$, where
$\phi_i$ is the spin at site $i$ and the first sum is over all
nearest-neighbor pairs of sites.  When restricted to spin values $\pm
a$, for any constant $a$, the first sum is equivalent to the energy in
the Ising model. For $\lambda>0, b<0 $ the second term can be viewed
as an energy contribution depending on how the spin-values deviate
from $\pm \frac{\sqrt{-b}}{\sqrt{2 \lambda}}$, as the summand is
minimzed for $\phi_i=\pm \frac{\sqrt{-b}}{\sqrt{2 \lambda}}$. Taking
$b=-2\lambda$ and letting $\lambda \rightarrow +\infty$ we get a
sequence of models with spin values increasingly concentrated around
$\pm 1$, i.e., the spin values of the Ising model.  Just like for the
Ising model this energy is unchanged by a global change of sign of
the spins.

Now, for the Ising model with $D=4$ the list of rigorous results is
relatively short but still quite powerful. The inequalities of Sokal
\cite{sokal:79} show that for $D \geq 4$ the specific heat follows its
mean field critical exponent, and for $D\geq 5$ it is bounded.  For
$D=4$ these inequalities do not prove that the specific heat is
bounded, only that it cannot diverge faster than $\ln(|K-K_c|)$.
Moreover the inequalities of \cite{aizenman,MR821310} show that for
$K\geq K_c$ the magnetisation $M$ is bounded as $c_1\sqrt{K-Kc}\leq M
\leq c_2 (\ln(K-K_c))^{3/2}\sqrt{K-K_c}$, which also means that the
magnetisation is continuous at $K_c$.  For $K\leq K_c$ it has recently
been proven~\cite{ADC21} that, properly rescaled, the magnetisation
converges to a Gaussian in the thermodynamic limit. However, the
finite size critical region for the 4D-model with cyclic boundary lies
in the interval $K>K_c$, see Ref.~\cite{lundow:11} for illustrations
of how the location of the critical region varies with dimension and
boundary condition, so this does cover all $L$-dependent finite size
effective critical points.  Non-rigorous results starting with
\cite{LK69} has predicted that for a class of models including the
Ising model the specific heat should diverge as $(\ln|K-K_c|)^{1/3}$,
but it has not been possible to turn this into a rigorous argument.
The model has also been studied via series expansions of the specific
heat and the susceptibility \cite{sykes:79, vohwinkel:92, oeis:20},
mainly leading to estimates for $K_c$. However the authors of
Ref.~\cite{MW06} attempts to estimate the exponent of $\ln|K-K_c|$
and cautiously note that this turns out to be difficult after getting
an estimate much larger than $1/3$.  Over the years, classical Monte
Carlo metods have also been applied \cite{PhysRevB.22.4481,
  Sanchez_Velasco_1987,PhysRevD.66.024008, lundow:09a} but the cost
for simulation in $D=4$ has kept the lattice sizes down, with
\cite{PhysRevD.66.024008} reaching $L=40$ and \cite{lundow:09a}
$L=60$. Most of these papers have estimated $K_c$ and simply concluded
that a $(\ln|K-K_c|)^{1/3}$-divergence is compatible with the sampled
data, but with no clear signal due to the limited range for $L$.
However, in Ref.~\cite{lundow:09a} we also took the microcanonical
ensemble into account and found that this favored a scenario where
the specific heat instead is bounded.  Later papers have also applied
numerical renormalisation techniques
\cite{PhysRevD.100.054510,Akiyama_2020}.

For the $\phi^4$-model the rigorous results are today much more
developed.  The same non-rigorous results as for the Ising model
\cite{LK69} applies here and predicts the same divergent specific
heat, as did later non-rigorous renormalisation arguments
\cite{WR73,PhysRevD.8.2418}.  The rigorous results from
\cite{sokal:79,aizenman,MR821310} also applies and proves that the
mean-field exponents are correct.  In 1989 Hara and Tasaki
\cite{MR892924, MR892925} finally rigorously proved that the predicted
logarithmic divergences are correct. Their results have since then
been extended and reproven by additional techniques, and
\cite{MR3269689, ADC21} both provide good overviews of what is now
known for this model.

Apart from the less well known Ref.~\cite{LK69} the advent of the
concept of universality in the 1970's led some authors --- it is not
clear if anyone can lay claim to be first --- to state that the Ising
model should have the exact same logarithmic divergence as the
$\phi^4$-model when $D=4$.  This claim is based on the fact that the
family of models with the same spatial dimension and same symmetry
group for the Hamiltonian, here given simply by sign-change for spins,
form the simplest candidate for a universality class.  Though Kadanoff
did already in Ref.~\cite{Kad76}[Page 18] point out that this is
merely the first approximation of the properties which define a
universality class by adding to the list of defining properties: {\it
  "Perhaps other criteria"}.  The belief in a simple universality
class was also strengthened by the rigorous proof in \cite{MR428998}
of the fact that by partitioning the 2-dimensional square lattice into
blocks and summing the spins inside each block to a block-spin value,
we can build a sequence of spin models which converge, in a certain
sense, to the 2-dimensional $\phi^4$-model. However, note that these
block-spins are by construction always bounded in value, according to
the block size, and the $\phi^4$-spins are unbounded.  So here the
exact definition of convergence is important and fluctuations in the
approximating block-spins are in some sense always smaller than those
in the limiting $\phi^4$-model.

Our aim in this paper is to investigate the critical behaviour of the
4-dimensional Ising model and, in addition to further sharpening
estimates for the location of the critical point, find clear evidence
for which universality class the model belongs to.  We have done this
by Monte Carlo simulation, keeping track of both microcanonical data
and the usual canonical ensemble data. We have used lattices of size
up to $L=256$, thus going far beyond earlier simulation studies. For
each lattice size we have run simulations at $150$--$300$
temperatures, and typically several hundred independent spin systems
of each size.  In the coming sections we first give definitions,
describe our sampling in more detail, and then proceed to analyze our
data, first in the microcanonical and then the canonical ensemble,
before finally coming to a discussion of our results.

\section{Definitions}
The underlying graph is the four-dimensional (4D) $L\times L\times
L\times L$ grid graph with periodic boundary conditions, i.e., the
Cartesian graph product of $4$ cycles of length $L$, thus having
$N=L^4$ vertices and $4L^4$ edges. On each vertex $i$ we place the
spin $s_i=\pm 1$ and let the Hamiltonian with interactions of unit
strength along the edges be $\mathscr{H}=\sum_{ij} s_i s_j$ where the
sum is taken over the edges $ij$. As usual the coupling $K=1/k_BT$ is
the dimensionless inverse temperature with $K_c$ the critical
coupling.

The magnetisation of a state $s$ is $M=\sum_i s_i$ (summing over the
vertices $i$) and the energy is $E=\sum_{ij} s_is_j$ (summing over the
edges $ij$), let also $m=M/N$ and $U=E/N$. With the partition function
$Z(K)=\sum_s e^{K E(s)}$ we can now define the standard quantities and
indicate how they can be measured. The internal energy is as usual
\begin{equation}\label{udef}
  \erg = \frac{1}{N}\frac{\partial\ln Z}{\partial K} =
  \frac{\mean{E}}{N}
\end{equation}
where $\mean{\cdots}$ is the thermal-equilibrium mean.  The specific
heat for a graph on $N$ vertices is defined as
\begin{equation}\label{cdef}
  \heat=\frac{1}{N} \frac{-\partial^2 \ln Z}{\partial T \partial K} =
  \frac{K^2}{N} \var{E}
\end{equation}
The energy excess kurtosis (or simply, kurtosis) is a ratio of
cumulants, or, a translated ratio of central moments
\begin{equation}\label{kurtdef}
  \ekurt = \frac{\partial^4\ln Z/\partial K^4}{\left(\partial^2\ln
    Z/\partial K^2\right)^2} = \frac{\mean{\left(E -
      \mean{E}\right)^4}}{\var{E}^{2}} - 3
\end{equation}

The data are collected in a micro-canonical fashion so that all
measurements are associated with the energy level $E$.  A crucial
quantity to measure is $R(E,\Delta E)$, defined as the probability at
energy $E$, that the energy changes by $\Delta E$ when a spin,
selected uniformly at random, is flipped.  Recall that for the spin at
vertex $i$ the energy changes by $\Delta E = -2s_i\sum_{ij} s_j$
(summing over the edges with one end in vertex $i$).

Let us briefly go through the micro-canonical details.  With each $K$
we associate an energy corresponding to the maximum term in the
partition function $Z(K)=\sum_E w(E)e^{KE}$, where $w(E)$ is the
number of states at energy $E$. Then
\begin{equation}\label{maxterm}
  w(E-\Delta E) e^{K(E-\Delta E)} < w(E) e^{K E} > w(E+\Delta E) e^{K(E+\Delta E)}
\end{equation}
leading to the relation, for $\Delta E>0$,
\begin{equation}
  \frac{1}{\Delta E}\ln\frac{w(E-\Delta E)}{w(E)}\le K \le
  \frac{1}{\Delta E}\ln\frac{w(E)}{w(E+\Delta E)}
\end{equation}
and we choose the average of the end-points as the value for $K$
\begin{equation}\label{kdef}
  \fK(U) = \frac{1}{2\Delta E}\ln\frac{w(E-\Delta E)}{w(E+\Delta E)}, \quad U=E/N
\end{equation}
With the (micro-canonical) entropy defined as $\fS(U)=(1/N)\ln w(E)$,
for $U=E/N$, its discrete derivative can now be written as
\begin{equation}\label{sder}
  \fS'(U)= \frac{\fS(U+\Delta E/N) - \fS(U-\Delta E/N)}{2\Delta E/N} = -\fK(U)
\end{equation}
Since $w$ and $R$ are related through the micro-canonical form of
detailed balance
\begin{equation}\label{detbal}
  w(E)\,R(E,\Delta E) = w(E+\Delta E)\,R(E+\Delta E, -\Delta E)
\end{equation}
we can now, using Eq.~\eqref{kdef}, give an alternative definition of
the $\fK$-function in terms of $R$
\begin{equation}\label{kfun}
  \fK(U) = \frac{1}{\Delta E}\ln\frac{R(E,-\Delta E)}{R(E,\Delta E)}, 
  \quad U=E/N, \, \Delta E>0
\end{equation}
where the approximation $R(E,\Delta E) \approx R(E+\Delta E,\Delta E)$
has been used implicitly~\cite{lundow:04a}.  Note that for 4D systems
a single spin-flip gives only $-16\le\Delta E \le 16$ and thus this
approximation is safe except for very small graphs. Note also that
there are four positive values of $\Delta E$ to choose from ($4, 8,
12, 16$) so we have in fact used a weighted average of the four
possible candidates for $\fK(E)$ leading to a small improvement in
data quality. This is in fact the only form of smoothing our data have
been subjected to.

From $\fK(U)$ we can now reconstruct the canonical energy distribution
for any $K$ as follows
\begin{equation}
  \label{predef}
  \Pr(E) = A \exp\left\{N \int_{u}^U\exp\left[K - \fK(x)\right] \dd x\right\}
\end{equation}
where $U=E/N$ and the constant $A$ is implicitly defined by $\sum_E
\Pr(E) = 1$. The lower bound $u$ in the integral is simply the
smallest $U$ for which we have data. Ideally one should measure $R$
for all $-4L^4\le E\le 4L^4$ but this is not practical for the larger
systems. One simply measures at a well-chosen range of $E$ such that
the energy distribution of Eq.~\eqref{predef} fits within, say, four
or five standard deviations from the end-points of the collected data
range. A detailed discussion of these rather technical considerations,
with worked examples, can be found in
Refs.~\cite{lundow:04a,lundow:09}.

Finally, the normalised coupling is denoted by $\varepsilon=(K-K_c)/K_c$ and
the rescaled coupling is $\kappa = L^2\varepsilon$. From time to time
we will also consider an alternative log-corrected form, $L^2(\ln
L)^{1/6}\varepsilon$.  A phenomenologically critical point is denoted
$K_c(L)$ or simply $K^*$ depending on context, analogously we write
$\varepsilon^*$ and $\kappa^*$. When a quantity depends on $L$ we will
subscript it, as in for example $\heat_L(K)$, and usually let
$\heat(K)$ refer to the asymptotic function (in the thermodynamic
limit) when $L\to\infty$.

\section{Ensemble equivalence}
Since the main evidence for our final conclusion comes from the
microcanonical ensemble, while the canonical ensemble is the one most
commonly used, we will first review the rigorous results on ensemble
equivalence.

Equivalence results have in the past been proven for various models,
varying assumptions, and different versions of equivalence. However,
more recently these results were proven in a rigorous and unified way
in Ref.~\cite{tou15}.  There it is proven that, assuming that the
thermodynamic limit of the model exists and is non-trivial, the
canonical and microcanonical ensembles are equivalent if and only if
the entropy $S(U)$ has a global supporting line at every point $U$ in
the support of $S$. Here a global supporting line at a point $U_0$
line is a tangent line for the graph at $S(U_0)$ which does not
intersect the graph at any other point.  In particular, such
supporting lines exist if $S$ is a strictly concave function. Note
that this theorem requires no other assumptions about the model or
exact behaviour at phase transitions.

If the ensembles are not equivalent it follows that $S(U)$ is not a
strictly concave function. This in turn means that the mode must have
meta-stable states for some temperature and a first-order phase
transition at that point.  The connection to first-order transitions
is nicely surveyed in Ref.~\cite{TOUCHETTE_2005}.

Looking at the Ising model in four dimensions we can quickly rule out
a first-order phase transition. Recall that at a first-order phase
transition the energy variance is unbounded, since the model jumps
between two states with macroscopically different energies $U_{-}$ and
$U_{+}$, so the only point where this could potentially happen is at
the usual critical point of our model. However, as already mentioned
the rigorous results of Refs.~\cite{aizenman,MR821310} show that the
magnetisation of our model is a continuous function, thereby ruling
out a first-order phase transition. So, for the 4-dimensional Ising
model we have ensemble equivalence.

We will now make this more quantitative.  For finite $N$ the free
energy $\free(K) = (1/N)\ln Z(K)$ satisfies $\free(K)\ge (1/N)
\ln(w(E)e^{KE})$ for all $E$, in particular for the maximum term at
$E$ associated with $K$, that is, $\free(K)\ge KU + S(U)$ where
$U=E/N$.  In the thermodynamic limit $N\rightarrow\infty$ $\free(K)$
is given by the Legendre-Fenchel transform of $-S(U)$, that is,
$\free(K)=\sup_U\{KU-(-S(U))\}$, since we have equivalence of the
canonical and microcanonical ensembles for the model.

To be precise, the condition of strict concavity means that if $S(U)$
is twice differentiable the ensembles are equivalent if
\begin{equation}\label{cvskd1}
 -\frac{(S'(U))^2}{S''(U)}\geq 0,
\end{equation}
with equality in at most one value of $U$.  In the thermodynamic limit
this is a condition on the specific heat, since
\begin{equation}\label{cvskd}
  \heat(K) = K^2 \frac{\partial\erg}{\partial K} =
  \frac{\fK(U)^2}{\partial \fK/\partial U} = \frac{\fK(U)^2}{\fK'(U)}= -\frac{(S'(U))^2}{S''(U)}
\end{equation}
Now let us assume that for $K$ close to $K_c$ and $U$ close to $U_c$ we have that  $\heat(K)=f(K_c-K)$ for some function $f$, which has a singularity of some form at 0. Now using Equation \ref{cvskd} and the fact that $\fK(U_c)=K_c$ this leads to
 $$\fK'(U)=\frac{(K_c+o(1))^2}{f(K_c-K)}.$$ Here we see that the
specific heat is divergent at $K_c$ if and only if $\fK'(U)\to 0$ when
$U\to U_c$.  The exact form of the divergence, e.g. power-law or
logarithmic, will only influence how quickly $\fK'(U)\to 0$ for $U$
close to $U_c$.  Similarly the specific heat will have a jump
discontinuity if and only if $\fK'(U)$ has one a well.

For a sequence of finite systems this in turn means that in order for
the maximum of the specific heat to diverge as $L$ increases, the
minimum value of $\fK_L'(U)$ must go to 0, as otherwise there would be
a strictly positive lower bound on the value of $\fK'(U)$

While these results demonstrate how the asymptotic, thermodynamical,
limits of the microcanonical and canonical ensembles connect do not
give us a simple connection between their finite-size approach to the
thermodynamical limits. For the canonical ensemble the finite-size
behaviour is a combination of the finite size effects for the
microcanonical density of states at individual energies, plus the fact
that the number of distinct energies grows with the system size, and
how strongly the exponential reweighting of the microcanonical states
concentrates the internal energy for given $K$ on a specific value of
$U$.  The latter can e.g. lead to an increase in the specific heat
even if there are no finite-size effects at all in the values of
$S(U)$ for those $U$ which correspond to existing values of $E$ for
finite $N$.  Over all, finite size effects for the canonical ensemble
are expected be more complex than for the micronanonical.

\section{Sampling and data analysis}
We have collected data for $L=6$, $8$, $10$, $12$, $16$, $20$, $24$,
$32$, $40$, $48$, $56$, $64$, $80$, $96$, $112$, $128$, $160$, $192$
and $256$ using standard Wolff-cluster updating~\cite{wolff:89} for
generating states in combination with the Mersenne-twister
random-number generator~\cite{matsumoto:98}.

The data are collected over an interval of energies covering all
points of phenomenological interest, such as maximum kurtosis and
specific heat. This is then a data window of particular interest where
we have the largest number of measurements per energy level. When
converting from micro-canonical to canonical form in this window we do
so at $\kappa$-steps of length $\approx 0.05$. Outside this window we
use larger step lengths, usually at most $0.25$. This typically means
that the various quantities are evaluated at $150$--$300$ different
temperatures depending on system size. For temperatures between these
steps we use standard third-order interpolation, thus effectively
giving us a continuum of temperatures.

The number of measurements per energy level vary considerably between
system sizes, for the smaller systems typically $10^4$--$10^6$, but
this number decreases quickly and for the largest systems it is only
$10$--$50$. The number of energy levels in the window of interest also
increases with $L$ so the total number of measurements is still large.
Keep in mind that the data are not smoothed at all (except for an
average taken in connection with Eq.~\eqref{kfun}) which means that
the $\fK$-function is quite noisy. However, the distribution resulting
from applying Eq.~\eqref{predef} to the $\fK$-function is very smooth
and well-behaved, since the noise is uncorrelated and hence mostly
canceled out by the integral. Once this distribution has been computed
it is of course easy to compute the weighted averages the canonical
quantities are made of.

This method produces very smooth canonical data, even though the
unsmoothed micro-canonical data appear noisy. Still, some mild noise
will show up in high- or low-temperature regions for higher-order
cumulants, especially for the largest systems. Inside the scaling
window such quantities usually are considerably more stable.

However, as mentioned in the discussion of ensemble equivalence the
error for a canonical quantity is a mixture of the errors for the
microcanonical data, meaning that we do not get a simple formula for
these errors.  Having said that, an indirect estimate of any error can
be obtained based on the Monte Carlo data gathered at a few
temperatures close to $K_c$ for some of the larger systems. For
example, for $L=256$ (and most of the larger $L$) we have roughly 2
million measurements at each of these temperatures.  From standard
bootstrap estimates of these data we find the error of the energy
$\erg$ to be less than $2\times 10^{-7}$ and for the specific heat
$\heat$ we estimate the error to be less than $0.003$.  Thus, for
these temperatures, we can use the raw Monte Carlo data for comparison
to our micro-canonical data.  Concretely we have found the largest
difference in the specific heat for $L=256$, to be 0.8\% of the
estimated value. The error for most temperatures and $L=96,\ldots,
256$ is typically in the range 0.1-0.6\%. For the internal energy the
largest error was $2\times 10^{-6}$, but more typically a magnitude
smaller, for $L\geq 80$. For smaller lattices both the specific heat
and internal energy have negligible errors. Hence we choose not to
print error bars in our figures.

We also expect that any error will show up as noise over an interval
of $K$, or that a sequence of system sizes will flush out any culprit
system that appears off in a finite-size scaling fit.  We are
particularly on the look-out for trends in the error, say if the
difference between the points and the fitted curve increases with $L$,
always a sign that the ansatz curve is not correctly chosen.  Often
there are higher-order corrections at play, so that the fitted curve
is only relevant for larger $L$. To guard against this we use the
standard technique of fitting the curve to a range of $L$ beginning at
$L_{\min}$, then repeat the fit by increasing $L_{\min}$. So, if the
ansatz is a curve of the form $A_0+\frac{A_1}{L^p}$ we can take the
median or mean of the $A_0$ and $A_1$ and use their respective
interquartile range or standard deviation as error estimates
(choosing, say, the largest).  To conclude, we check against trends in
$L$ and often estimate error bars in several different ways. This is a
more qualitative and, we think, relevant approach to data analysis
than simply using reduced $\chi^2$-estimates of the error, especially
when non-linear data are involved~\cite{andrae:10}.

\section{Micro-canonical quantities}\label{sec:micro}
\begin{figure}
  \includegraphics[width=0.483\textwidth]{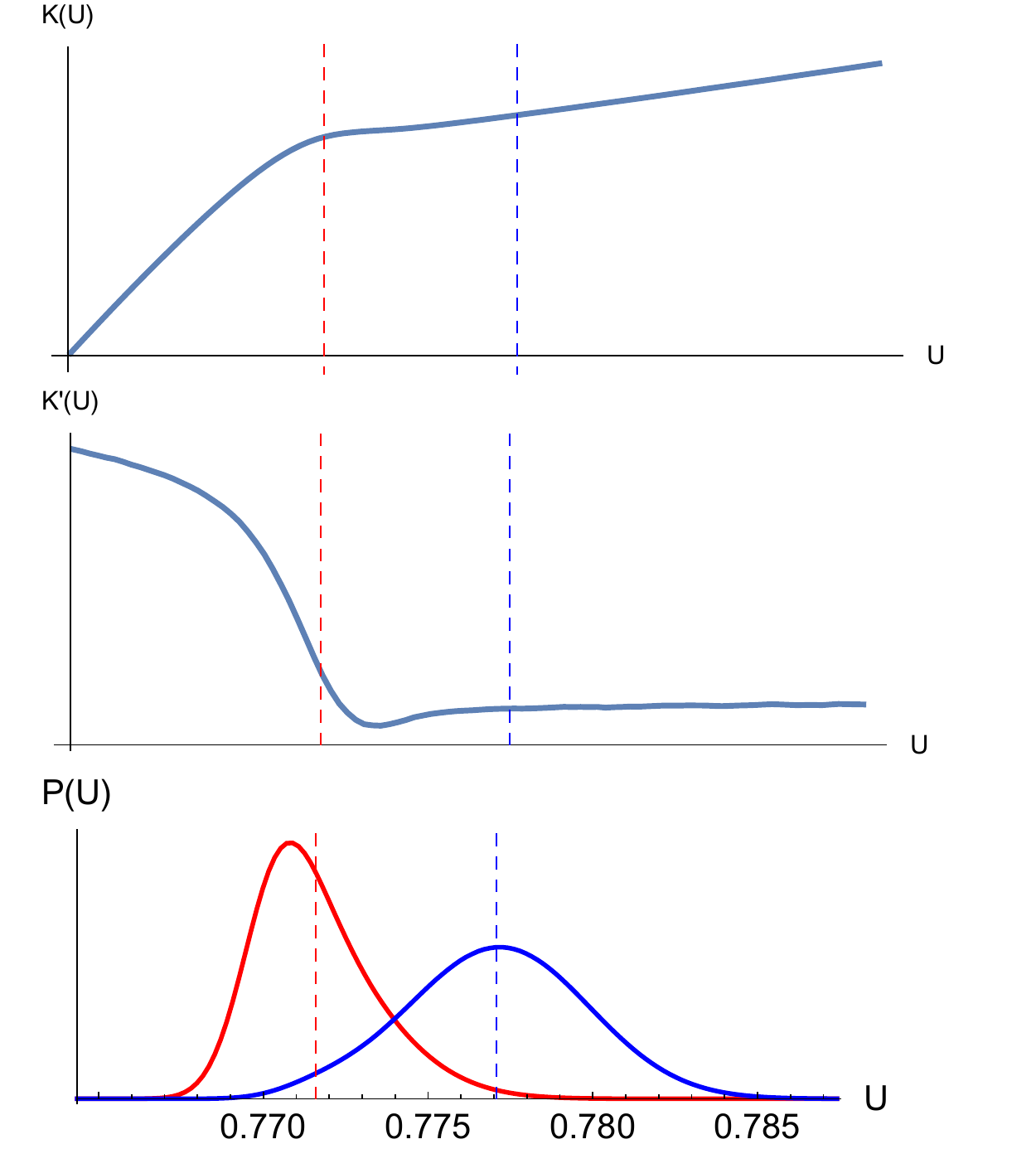}
  \caption{\label{fig:kkdd}(Colour on-line) $\fK_L(U)$ (top),
    $\fK_L'(U)$ (middle) and $\Pr(U)$ (bottom) plotted versus $U$ for
    $L=64$. All three figures cover the same interval of $U$. The
    bottom figure shows the distribution of energies for $K=K_c$
    (left) and for $K^*$ (right) giving the maximum specific heat.
    The dashed lines show the mean energy, $\erg(K)$, for the
    respective distribution (or $K$). This picture is representative
    for all $L\ge 6$ regarding the shape of the distributions and
    their location in relation to the minimum in $\fK_L'(U)$.  }
\end{figure}

The micro-canonical density of states is of course fundamental to
the model and determines the behaviour of the canonical
distribution. Fig.~\ref{fig:kkdd} demonstrates this in the case of
$L=64$, showing both $\fK_L(U)$ and its derivative $\fK_L'(U)$,
together with the energy distributions for two temperatures of
interest, $K_c$ and at the $K_c(L)$ corresponding to the maximum
specific heat. This figure is quite representative for all $L\ge
6$. For example, the distribution at $K_c$ is sharply skewed to the
left, somewhat off-center from the minimum $\fK'(U)$.  At $K_c(L)$ on
the other hand, the distribution has a more symmetric look. All this
is of course governed by the $\fK$-curve. Other phenomenological
critical points occur at other places giving rise to distributions of
other shapes. For the purpose of this plot the $\fK_L$- and
$\fK_L'$-curves have here been smoothed, using a standard moving-window 
average of width corresponding to half the standard deviation
of the energy at $K_c(L)$.

\begin{figure}
  \includegraphics[width=0.483\textwidth]{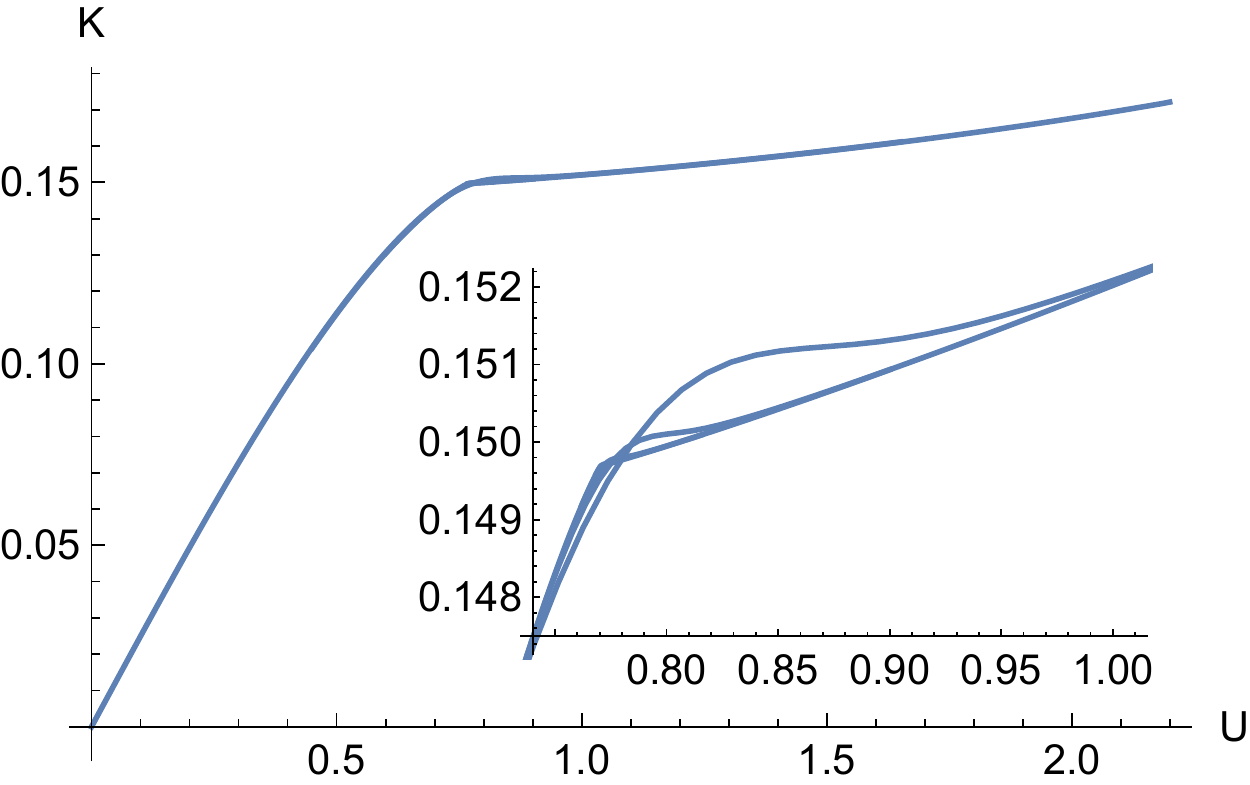}
  \caption{\label{fig:k}(Colour on-line) Function $\fK_L(U)$ plotted
    versus $U$ for $L=8$, $16$, $32$, $64$, $128$. Inset shows a
    zoomed in plot. The curves are smoothed.}
\end{figure}

\begin{figure}
  \includegraphics[width=0.483\textwidth]{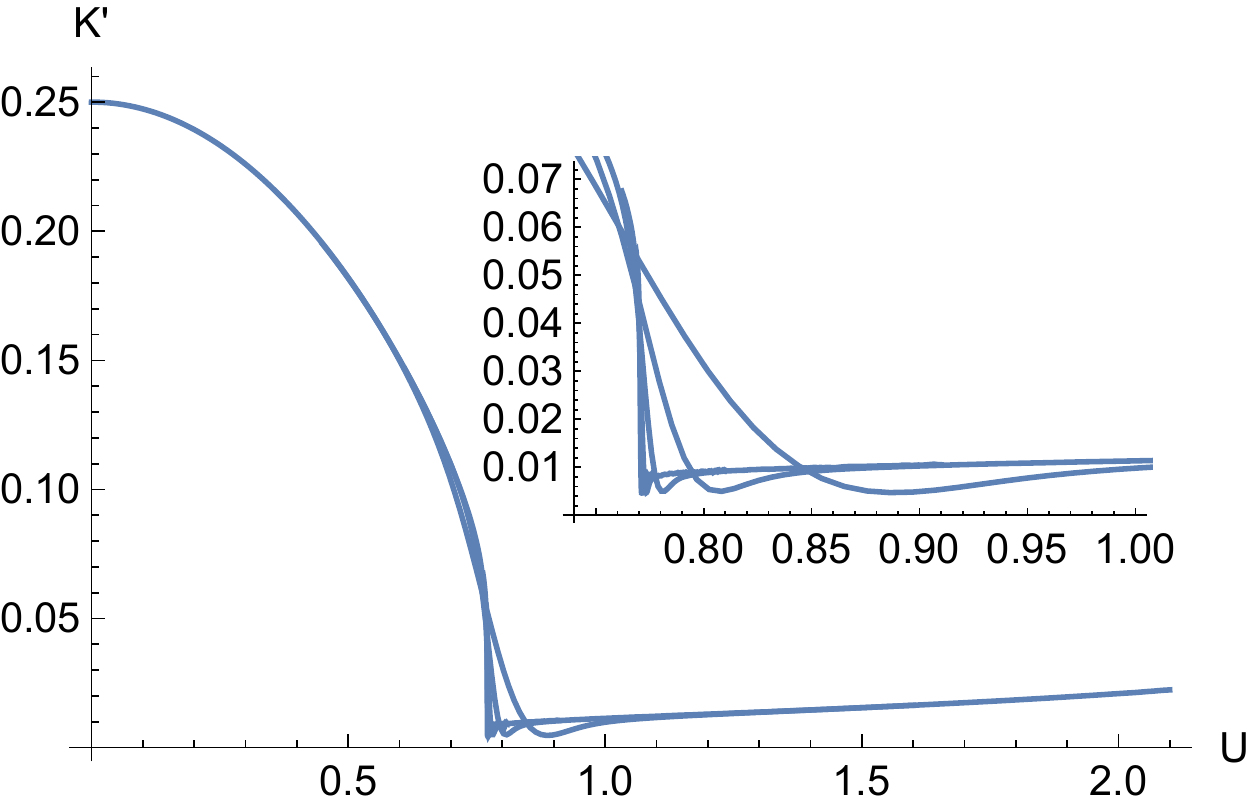}
  \caption{\label{fig:kd}(Colour on-line) Function $\fK_L'(U)$ plotted
    versus $U$ for $L=8$, $16$, $32$, $64$, $128$. Inset shows a
    zoomed in plot. The curves are smoothed.}
\end{figure}

\begin{figure}
  \includegraphics[width=0.483\textwidth]{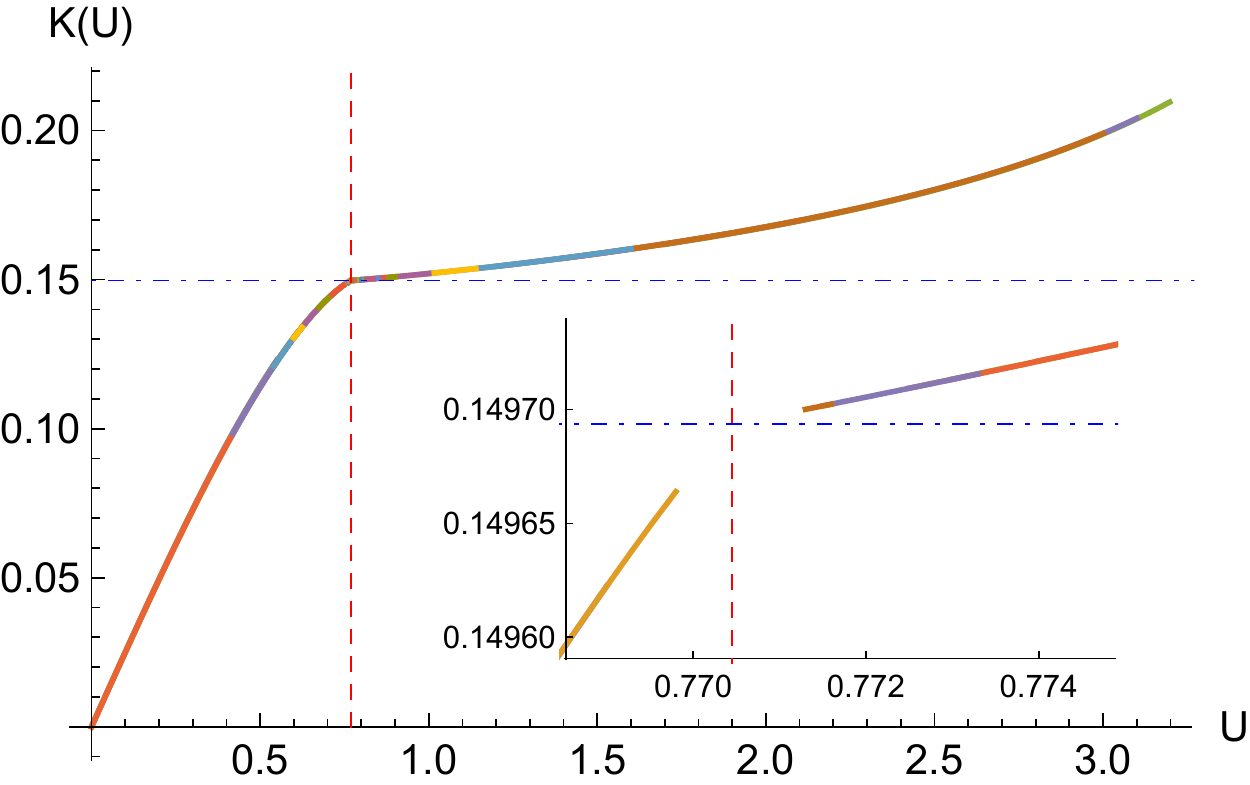}
  \caption{\label{fig:klim}(Colour on-line) Limit
    curve $\fK(U)$ plotted versus $U$. Pieced together from smoothed
    $\fK_L(U)$ for $6\le L\le 256$ (different colours). The dashed
    vertical line is at $U_c$ and the horisontal dot-dashed line at
    $K_c$. Inset shows a zoomed in plot.}
\end{figure}

\begin{figure}
  \includegraphics[width=0.483\textwidth]{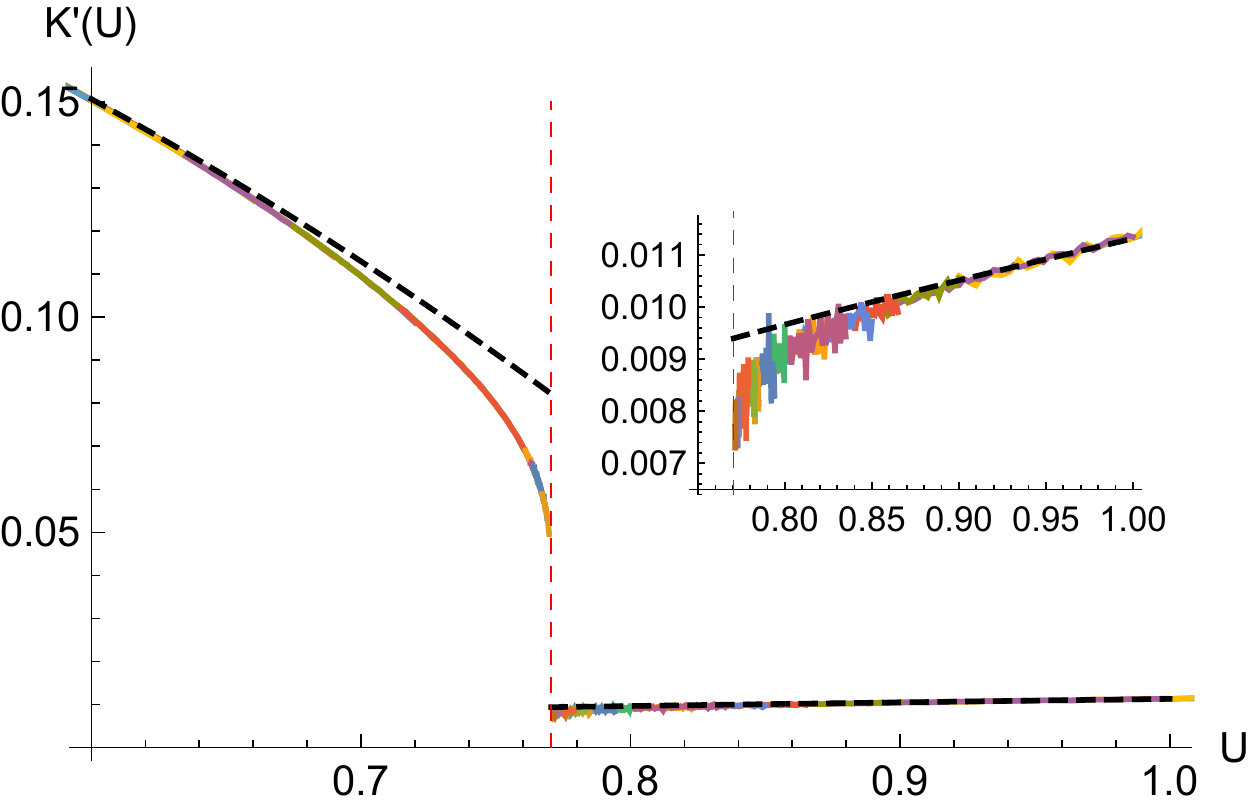}
  \caption{\label{fig:kdlim}(Colour on-line) Limit curve $\fK'(U)$
    plotted versus $U$. Pieced together from smoothed $\fK_L'(U)$ for
    $6\le L\le 256$ (different colours). The dashed vertical line is
    at $U_c$. The dashed black curve is obtained from series
    expansions. Inset shows a zoomed in plot of the low-temperature
    case.}
\end{figure}

In Figs.~\ref{fig:k} and \ref{fig:kd} we plot $\fK_L(U)$ and
$\fK_L'(U)$ respectively for a range of $L$ to show how they evolve to
a limit curve. Piecing together a sequence of $\fK$- and
$\fK'$-curves, according to where their values agree with those for
larger $L$, gives an approximation of their asymptotic limit and are
shown in Figs.~\ref{fig:klim} and \ref{fig:kdlim}.  Note that, as seen
in the inset of Fig.~\ref{fig:klim} this approximation will not give a
curve very close to the critical point, since the data from the
largest value of $L$ cannot be compared to a larger size.  In the
second figure we have also included data from Pad\'e-approximants
based on the high- and low-temperature series expansions of the free
energy~\cite{sykes:79,vohwinkel:92,oeis:20}.

\begin{figure}
  \includegraphics[width=0.483\textwidth]{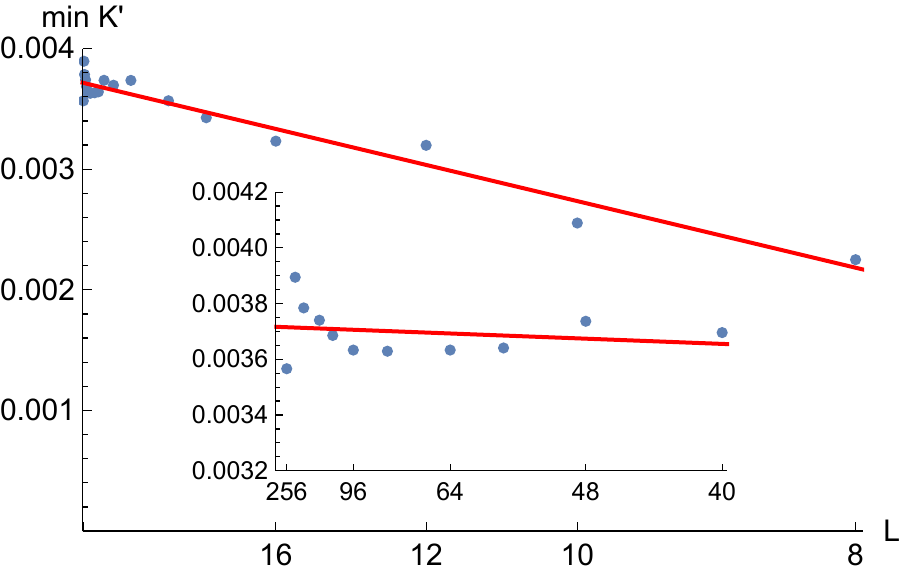}
  \caption{\label{fig:kdmin}(Colour on-line) Minimum $\fK_L'(U)$
    plotted versus $1/L^2$ for $L=8,\ldots, 256$. The fitted red line
    is $y=0.00371(8)-0.098(1)x$ where $x=1/L^2$. The inset shows a
    zoom-in for $L=40,\ldots,256$.}
\end{figure}

Of particular interest here is of course the minimum value of
$\fK_L'(U)$ and these are shown in Fig.~\ref{fig:kdmin}. We have in
this plot attempted a simple scaling rule for these points,
$y=A_0+A_1/L^2$ fitted to $L\ge 8$, which appear largely correct,
despite the presence of some noise.  We estimate a limit value of
$A_0=0.00371(8)$ and the slope $A_1=-0.098(1)$. The error bar for
$A_0$ is here the mean difference between the line and the points and
for $A_1$ the standard deviation of the fitted slopes when deleting
one point.  Here we work under the assumption that the model does not
have a finite-size effect so large that the minimum $\fK'(U)$ begins to
decrease for some $L > 256$.

In our earlier study~\cite{lundow:09a} we only had data for $L\le 80$
and used $A_1/L$ as correction term but with these new data that
correction term does not fit as well as $A_1/L^2$. However, the result
is qualitatively the same: the minimum value is increasing with $L$
and, more importantly, the limit is distinctly positive.

As a  technical  point we note that care must be taken when finding the minimum
$\fK'_L$.  We have fitted a $7$th degree polynomial to the unsmoothed
$\fK_L(U)$-data over a $U$-interval where it preserves the shape
and profile of the $\fK_L'(U)$-curves for different $L$. This interval
is of course easily recognized by comparing the derivative of the
polynomial to the smoothed $\fK_L'$-curves.

Recall that in the thermodynamic limit Eq.~\eqref{cvskd} derives the
maximum specific heat from the minimum value of the
$\fK_L'(U)$-curve. However, here the order of the limits is crucial;
the limit of the minima for finite $L$ is not necessarily equal to the
minimum of the limiting curve. Such examples are known already from
the 2D Ising model.  If we look at the sequence of curves in
Fig.~\ref{fig:kd} we see that to the right of the critical point the
curves step by step agree on a curve which seems to lie noticeably
higher than the sequence of minima.  From the rough asymptotic curve
given in Fig.~\ref{fig:kdlim} we see that the curve does bend down
close to the critical point and we can give 0.008 as a, very, rough
upper bound on the asymptotic minimum, while $A_0$ of course provides
a lower bound.

Following Eq.~\eqref{cvskd} this would imply that the limit maximum
specific heat, given by $K_c^2/(\min \fK') $ is bounded by $
0.1496^2/0.008\approx 2.8 \lesssim \heat_{\infty} \lesssim
0.1496^2/0.0037 \approx 6.03$.  Since these two bounds are based
strictly on the values to the right of $U_c$ they are also bounds on
the limit of $\heat_{\infty}$ from the low-temperature side.  Using
the rough upper bound of $0.05$ for the left side limit of $\fK'$ we
find $ 0.1496^2/0.05 \approx 0.448 $ to be a lower bound on the limit
of $\heat_{\infty}$ from the high-temperature side, while the global
bound provides an upper bound.

\begin{figure}
  \includegraphics[width=0.483\textwidth]{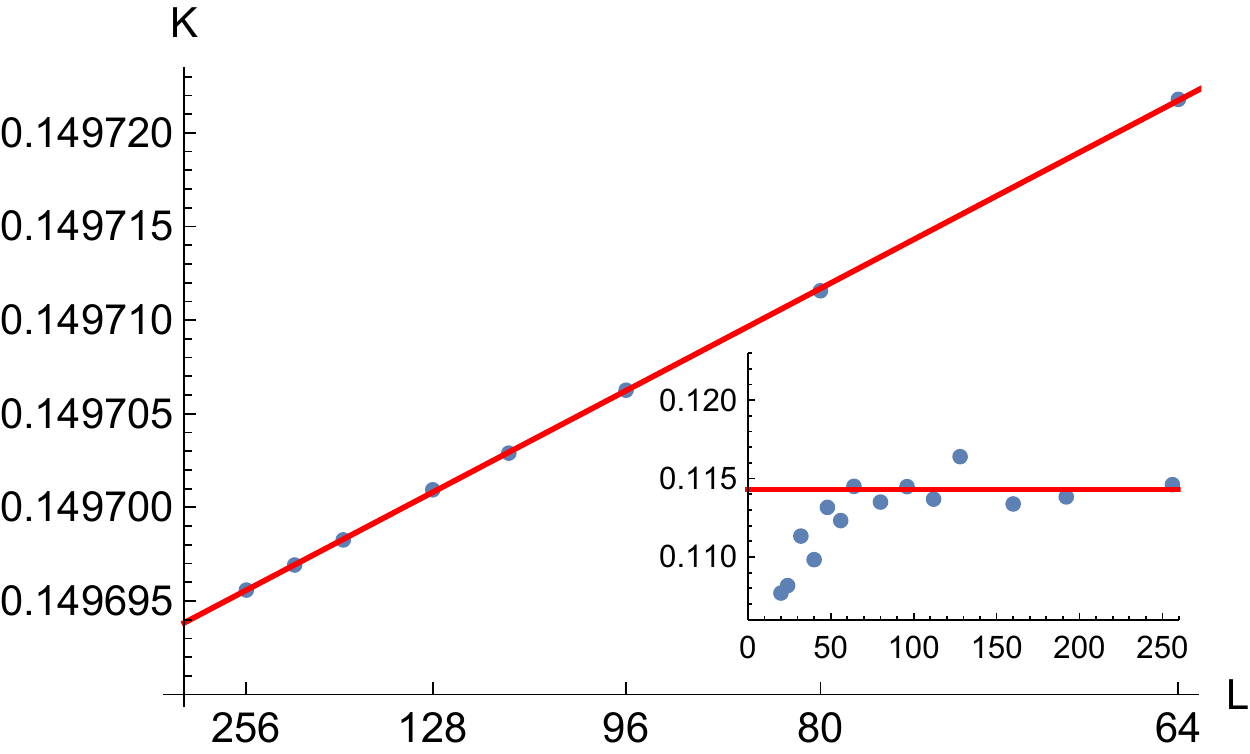}
  \caption{\label{fig:kkdmin}(Colour on-line) The value of $\fK_L$ at
    minimum $\fK_L'(U)$, or $K_c(L)$, plotted versus $1/L^2$ for
    $L=64$, $80$, $96$, $112$, $128$, $160$, $192$, $256$.  The red
    line, fitted to these points, is $y=0.14969383(6)+0.1143x$ where
    $x=1/L^2$.  The inset shows $(K_c(L)-0.14969383)L^2$ versus $L$
    for $L\ge 16$ and the constant line $y=0.1143$.}
\end{figure}

It is of some interest to also consider $K_c(L)$, defined as the value
of $\fK$ at this $\fK'$-minimum, see Fig.~\ref{fig:kkdmin}. We find
$K_c(L)=0.14969383(6) + 0.1143(2)/L^2$ (fitting to $L\ge 64$) with
error bars obtained as above; mean distance between points and line,
and, slope variation when deleting each point in turn. Higher-order
corrections become relevant for $L<64$. As we will see later this
estimate of $K_c$ is consistent with what we will find  from the
canonical quantities.

\begin{figure}
  \includegraphics[width=0.483\textwidth]{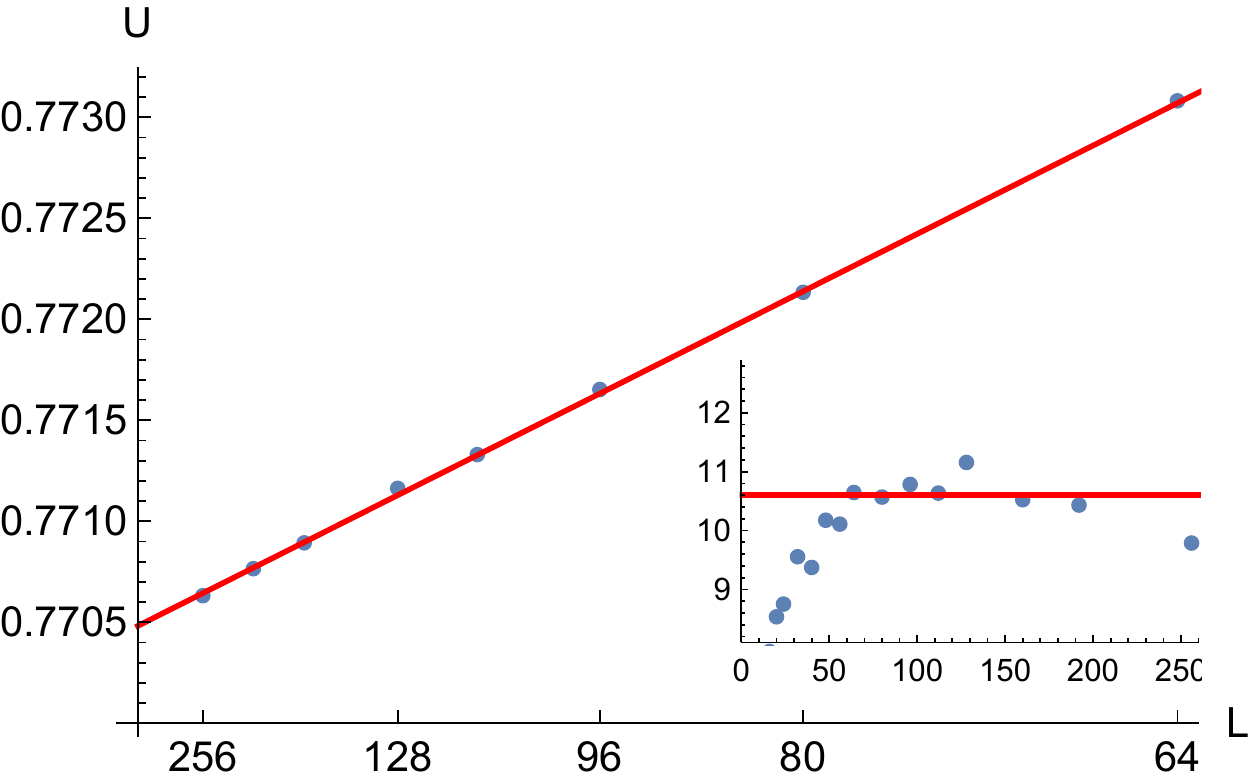}
  \caption{\label{fig:ukdmin}(Colour on-line) The value $U_c(L)$
    giving minimum $\fK_L'(U)$ plotted versus $1/L^2$ for $L=64$,
    $80$, $96$, $112$, $128$, $160$, $192$, $256$.  The red line,
    fitted to these points, is $y=0.77048(1)+10.60x$ where
    $x=1/L^2$. Inset shows $(U_c(L)-0.77048)L^2$ versus $L$ for $L\ge
    16$ and the constant line $y=10.6$.}
\end{figure}

Finally, the location $U_L$ of the minimum $\fK_L'(U)$ is shown in
Fig.~\ref{fig:ukdmin}. We find $U_L=0.77048(1)+10.60(4)/L^2$, obtained
in the same way as the scaling for $K_c(L)$ above, thus giving us an
estimate of $U_c$. Note that the inset of the figure shows that
$L=256$ deviates slightly from the line, but this $L$ also has more
noise than the smaller cases.

\section{Canonical properties: Finite-size scaling for the specific heat}\label{finscan}
We now turn to the properties of the model in the better known
canonical ensemble.  We will first test two scenarios for the
finite-size scaling of the specific heat at, or near, $K_c$.  These
are in turn based on two distinct asymptotic behaviours for the
specific heat, either it is bounded as in the pure mean-field model,
or it has a poly-log divergence of the same type as the
$\phi^4$-model. A challenge for earlier studies has been to
distinguish between these distinct asymptotics due to the slow growth
in the second scenario.

\begin{figure}
  \includegraphics[width=0.483\textwidth]{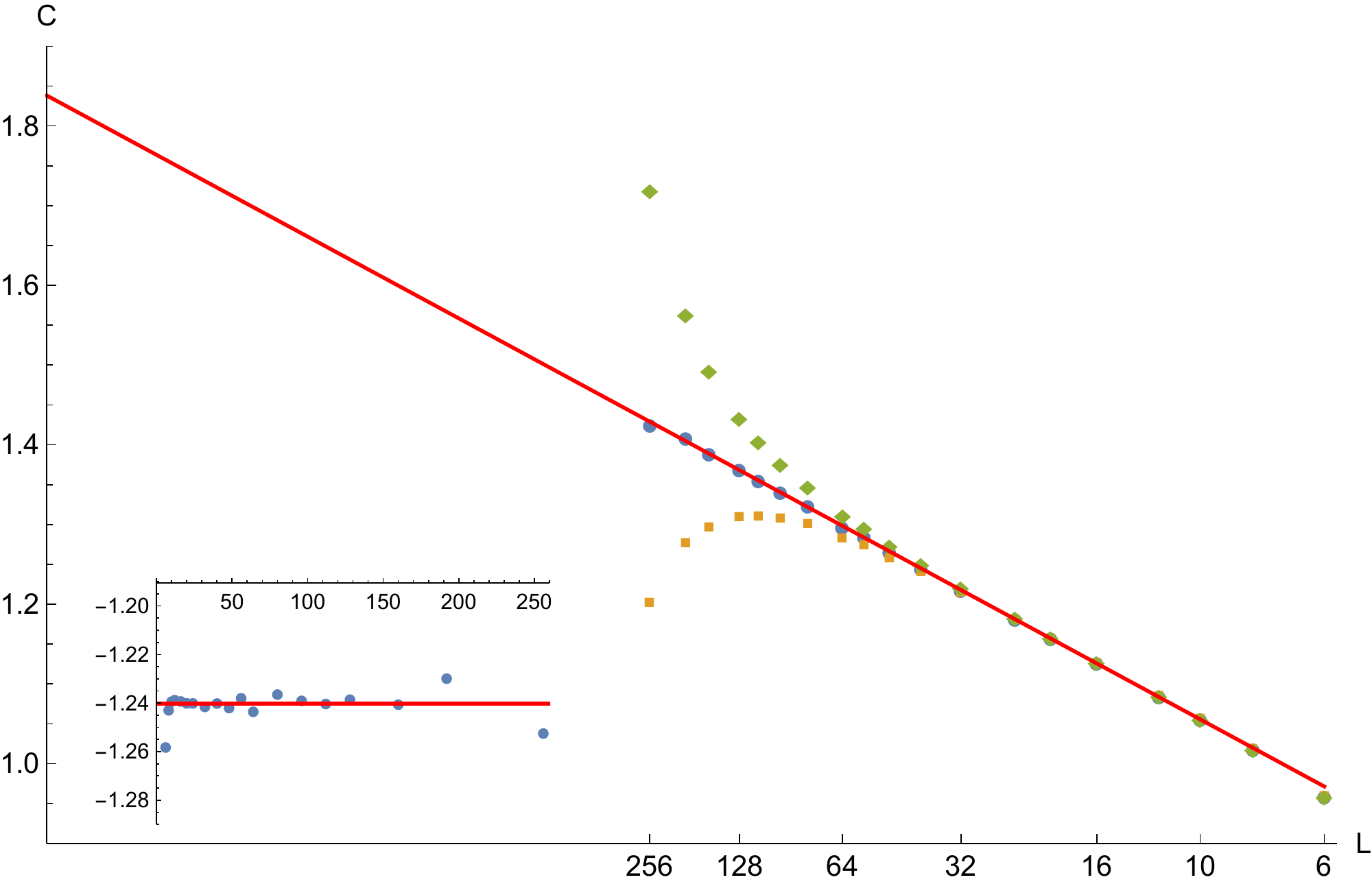}
  \caption{\label{fig:ckc1}(Colour on-line) Specific heat
    $\heat_L(K_c)$ versus $1/L^{1/5}$ for $L=6, 8,\ldots,256$ and the
    fitted line $y=1.838(3)-1.24(1)x$ where $x=1/L^{1/5}$. Here we use
    $K_c=0.14969378$. The other points are $\heat_L$ at $K_c+5\times
    10^{-7}$ (diamonds, bending upwards) and $K_c-5\times 10^{-7}$
    (squares, bending downwards). Inset shows correction to scaling
    $(\heat_L(K_c)-1.838)L^{1/5}$ versus $L$ and the constant line
    $y=-1.24$.}
\end{figure}

In the first scenario, where we assume that
the specific heat is bounded, we use a very simple scaling rule:
$\heat_L(K_c) = A_0+A_1/L^a$.  This turns out to give a good
first estimate of $K_c$ as a bonus. In Fig.~\ref{fig:ckc1} we show this
for $a=1/5$ using $K_c=0.14969378$. The points that bend upwards
and downwards shows the same sequence for $K_c\pm 5\times 10^{-7}$.
In fact, this effect is clear already for changes in $K_c$ as small as
$1.5\times 10^{-7}$. This sensitvity is obtained for a wide range of
$a<1$. Thus, so far we have $K_c=0.14969378(15)$.

However, the corrections to scaling $(\heat_L(K_c)-A_0)L^a$ can
give us a clearer picture as to the quality of the fitted
scaling. Plausible fits to the points (not shown here) can be made for
$a=1/6$ and $1/4$ (if we stay with the simple rationals) only
with differently signed corrections for small $L$. Outside this range
the corrections look less convincing. We find that the best fit is
found for $a=1/5$ using $K_c=0.14969378(1)$ where the error bar
indicates the interval where the corrections look similar to the inset
plot of Fig.~\ref{fig:ckc1}. Note that there are only very weak
corrections for small $L$ and some larger deviations for $L=192$ and
$256$. Fitting on $L_{\min}\le L\le 256$ for $20\le L_{\min}\le 64$
gives the asymptotic specific heat at $K_c$ being $A_0=1.837(3)$ and
the correction coefficient $A_1=-1.24(1)$.  The $A_0$-value
is of course highly dependent of $a$. Choosing $a=1/6$
gives $A_0\approx 1.94$ and $a=1/4$ gives $A_0\approx 1.74$ when
fitting on $L\ge 64$.

Let us also here mention that the global maximum of the specific heat 
  is excellently fitted by $\max \heat_L = 4.378(3) - 3.472(9) /
L^{1/5}$ with error bars from fitting over $L$-ranges with $20\le
L_{\min}\le 64$.

The second scenario, based on  calculations for
the $\phi^4$-model~\cite{rudnick:85a,lai:90,kenna:04}, is
\begin{equation}\label{crule1}
  \heat_L(K_c) =B_0+B_1(\ln L)^{1/3}+B_2\frac{\ln\ln L}{(\ln L)^{2/3}}
\end{equation}

\begin{figure}
  \includegraphics[width=0.483\textwidth]{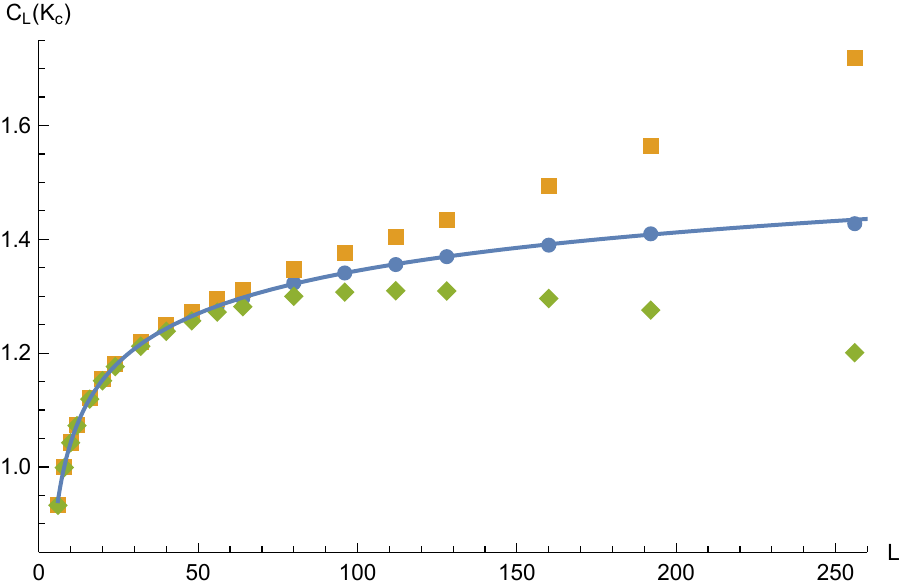}
  \caption{\label{fig:ckc2}(Colour on-line) Specific heat
    $\heat_L(K_c)$ versus $L$ for $L=6, 8,\ldots,256$ and the fitted
    curve $y=-0.15+0.849 (\ln L)^{1/3} + 0.15\frac{\ln \ln L}{(\ln
      L)^{2/3}}$ for the points at $K_c=0.149693785$. The other points
    are $\heat_L$ at $K_c+5\times 10^{-7}$ (orange squares, bending
    upwards) and $K_c-5\times 10^{-7}$ (green diamonds, bending
    downwards).}
\end{figure}

In Fig.~\ref{fig:ckc2} we show $\heat_L(K_c)$ versus $L$ with
$K_c=0.149693785$. To these points we fit Eq.~\eqref{crule1}, finding
$B_0=-0.15(1)$, $B_1=0.849(1)$ and $B_2=0.15(1)$ (error bars from
different point ranges). The two other point sets in the figure shows
the values for $K_c\pm 5\times 10^{-7}$, thus clearly giving an
interval for $K_c$. In fact, the effect is visible for $\pm 1\times
10^{-7}$.  The error in the fit is less than $0.001$ for $8\le L\le
192$. Hence, the second scenario, having the same number of parameters
as the first, also results in a good fit.



In the next section we make use of the two scaling scenarios to test their
respective scenario for the thermodynamic limit of the model.

\section{The asymptotic specific heat: Two scenarios}\label{2scen}

We will now test the two main scenarios for the behaviour of the
specific heat close to $K_c$ in the thermodynamic limit: {\bf a)}
that, like for higher dimensions $d\ge 5$~\cite{lundow:15}, it follows
mean-field behaviour: taking a left-limit value for $\varepsilon\to
0^-$, another value for $\varepsilon=0$, and a right-limit value for
$\varepsilon\to 0^+$, with some singular exponents $\theta^-$ and
$\theta^+$ guiding the behaviour for small values of $|\varepsilon|$,
or, {\bf b)} that it behaves as the
$\phi^4$-model~\cite{lai:90,kenna:04}, i.e., $C(\varepsilon)\sim
(-\ln(|\varepsilon|))^{1/3}$.

\subsection{The high-temperature range}
\begin{figure}
  \includegraphics[width=0.483\textwidth]{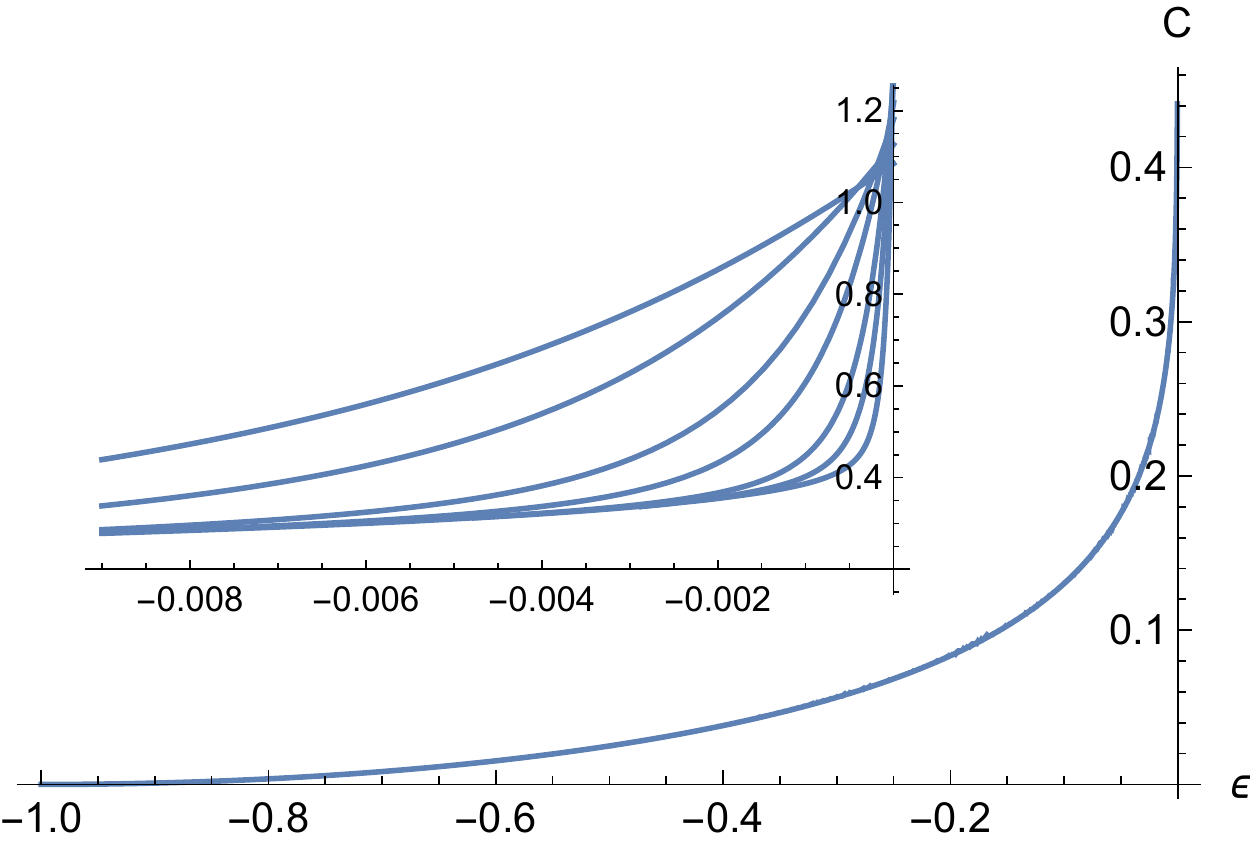}
  \caption{\label{fig:ceps1}(Colour on-line) Asymptotic specific heat
    $\heat(\varepsilon)$ versus $\varepsilon$ for high temperatures. Inset
    shows $\heat_L$ versus $\varepsilon$ for $L=6$, $8$, $12$, $16$,
    $24$, $32$, $48$, $64$, $96$. Asymptotic curve $\heat_{\infty}$
    built from piecing together $\heat_L$ for $6\le L\le 192$.}
\end{figure}

In order to test the scenarios we will first construct an asymptotic
($L\to\infty$) $\heat$-curve. This can be obtained by piecing together
$\heat_L$-data, for a sequence of $L$, proceeding like in
Ref.~\cite{lundow:15}.  In more detail, $\heat_6(K)$ and $\heat_8(K)$
are effectively equal (modulo noise) for $K\le 0.11$, determined by
ocular inspection with some safety margin, so $\heat_6(K)$ can now be
treated as the asymptotic $\heat(K)$-curve when $K\le 0.11$. Next,
$\heat_8(K)$ and $\heat_{10}(K)$ are effectively equal for $K\le
0.125$, so $\heat_8(K)$ can be treated as $\heat(K)$ for $K\le 0.125$,
etc. Continuing like this all the way up to $\heat_{160}(K)$ for $K\le
0.14965$ (our $\heat_L$-data for $L=192$ and $L=256$ unfortunately do
not overlap those for $L=160$) we have built up a limit
$\heat(K)$-curve for $0\le K\le 0.14965$, or preferably,
$\heat(\varepsilon)$ for $-1\le\varepsilon\lesssim-0.00029$. In
Fig.~\ref{fig:ceps1} we show the resulting asymptotic curve
$\heat(\varepsilon)$ together with an inset demonstrating how some of
the $\heat_L$-curves overlap.

\begin{figure}
  \includegraphics[width=0.483\textwidth]{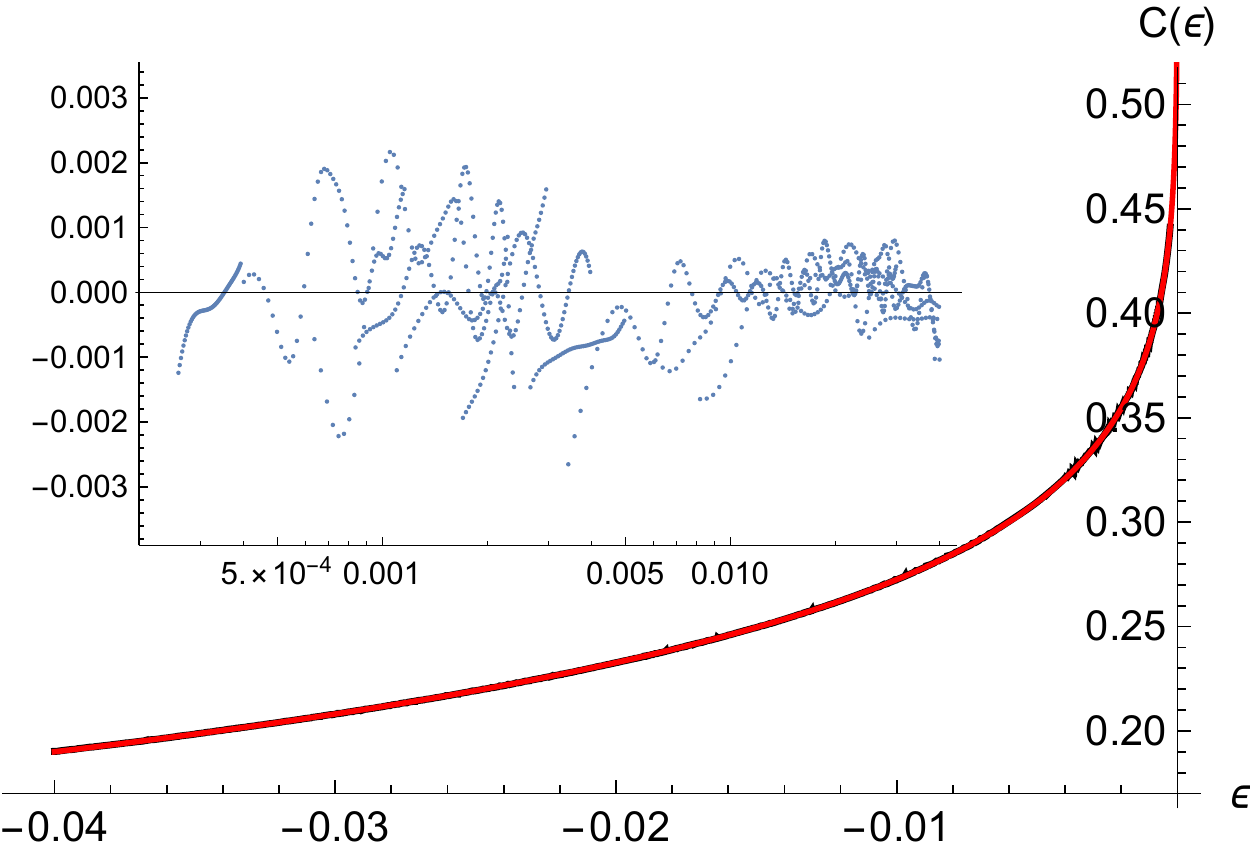}
  \caption{\label{fig:cfit1}(Colour on-line) Asymptotic specific heat
    $\heat(\varepsilon)$ and the fitted curve
    $y=0.807-0.861(-\varepsilon)^{0.1035}$ (indistinguishable) versus
    $\varepsilon$. Inset shows relative error $(y-\heat)/\heat$ versus
    $\ln(-\varepsilon)$.}
\end{figure}

\begin{figure}
  \includegraphics[width=0.483\textwidth]{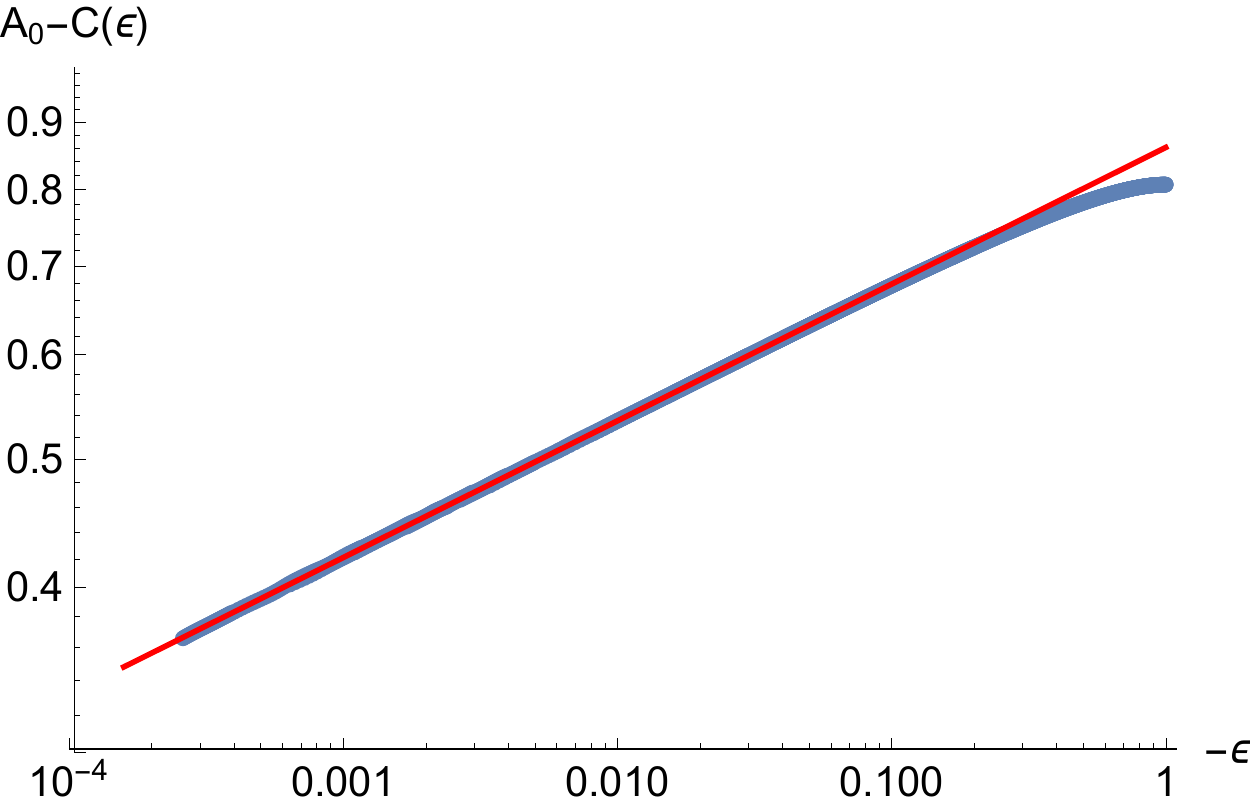}
  \caption{\label{fig:clog1}(Colour on-line) Log-log plot showing
    $0.807-\heat(\varepsilon)$ versus $-\varepsilon$. The red line is
    $y=\ln(0.861)+0.1035x$ where $x=\ln(-\varepsilon)$.}
\end{figure}

Fitting the asymptotic curve to a simple formula $y=A_0+A_1
(-\varepsilon)^{\theta^-}$ over $-0.04<\varepsilon$ gives
$\theta^-=0.1035(5)$ with $A_0=0.807(1)$ and
$A_1=-0.861(1)$. Adjusting the lower bound between $-0.02$ and $-0.05$
gives parameters varying only very slightly, thus providing us with
quite narrow error bars.  The resulting curve is shown in
Fig.~\ref{fig:cfit1}.  As the inset plot shows the relative error of
such a fit is small and shows no clear trend.  In Fig.~\ref{fig:clog1}
we show $0.807-\heat(\varepsilon)$ versus $-\varepsilon$ in a log-log
plot together with the line $y=\ln(0.861)+0.1035x$ to demonstrate its
linear behaviour.

Of course, if we narrow the fitted interval still further the
parameters will change, but only very little. For example, with
$-0.01<\varepsilon$ the curve becomes
$y=0.821-0.870(\-\varepsilon)^{0.100}$, but the relative error will
then grow larger at $\varepsilon=-0.04$. On a more global scale the
specific heat should then take an almost linear behaviour when plotting
$A_0-\heat(\varepsilon)$ versus $-\varepsilon$ in a log-log plot, see
Fig.~\ref{fig:clog1}. Of course, choosing different values of $A_0$
will change the picture, becoming visibly non-linear for $A_0<0.7$ and
$A_0>1.1$.

Having established the stability of the first scenario over a wide
interval of negative values for $\varepsilon$ let us now try the
second scenario.  Here we will fit $y=A_0 + A_1
(-\ln(-\varepsilon))^{1/3}$. However, this formula fails to fit our
data when the lower bound is less than $-0.01$ so we can not use the
much wider interval from the previous scenario. For example, we obtain
$y=-0.513+0.472(-\ln(-\varepsilon))^{1/3}$ when fitted to
$-0.01<\varepsilon$. If we include a correction term, see
Ref.~\cite{kenna:04}, we might hope to obtain a reasonably good fit
over a wider interval;
\begin{equation}\label{yfit2}
  y = A_0 + A_1 (-\ln(-\varepsilon))^{1/3} + 
  A_2 \frac{\ln(-\ln(-\varepsilon))}{(-\ln(-\varepsilon))^{2/3}}
\end{equation}
Note that we have taken the liberty to include a constant term $A_0$
as a catch-all term for the weaker corrections. Fitting to
$-0.04<\varepsilon$ we obtain $A_0=0.41(1)$, $A_1=0.457(1)$,
$A_2=0.14(2)$ (error bar from changing the lower bound between
$-0.025$ and $-0.045$).

\begin{figure}
  \includegraphics[width=0.483\textwidth]{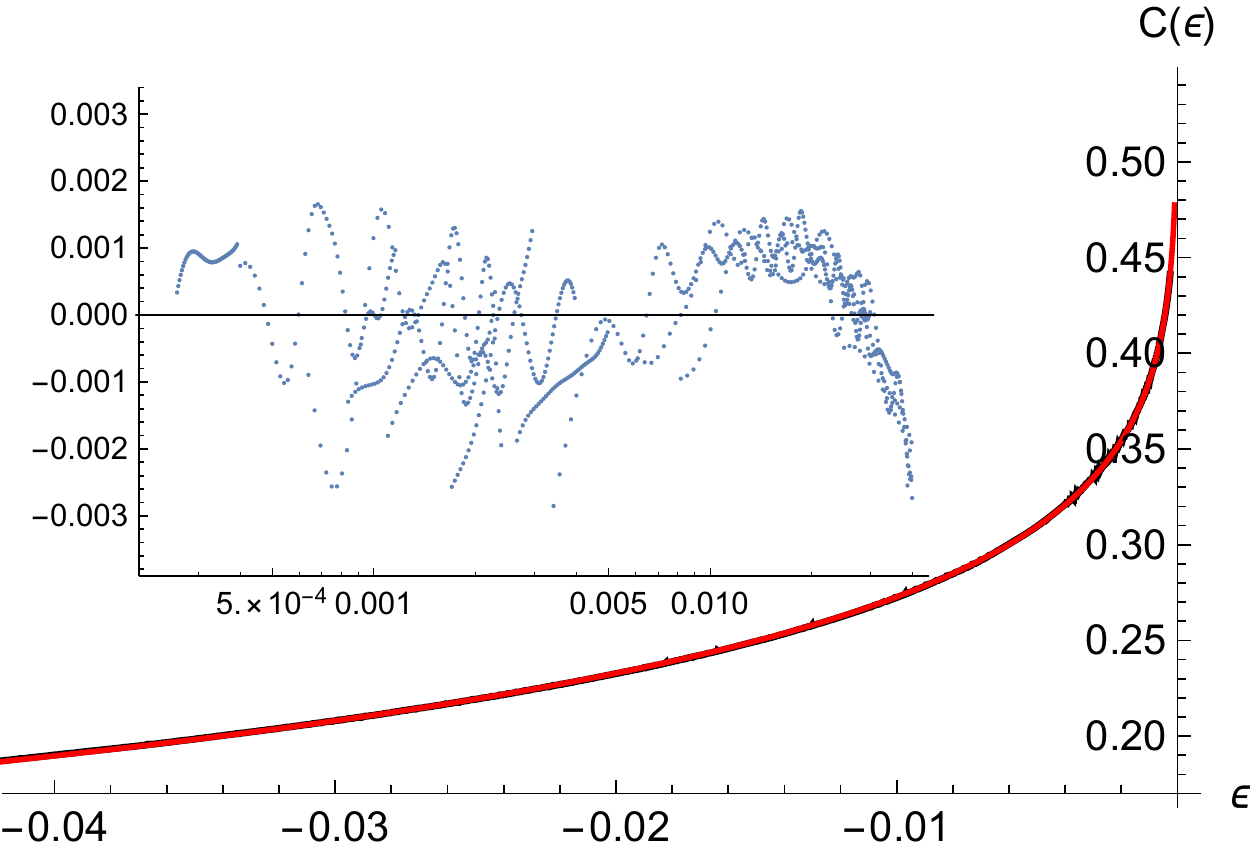}
  \caption{\label{fig:cfit2}(Colour on-line) Asymptotic specific heat
    $\heat(\varepsilon)$ and a fitted curve $y$ of Eq.~\eqref{yfit2}
    (indistinguishable) versus $\varepsilon$. Inset shows relative
    error $(y-\heat)/\heat$ versus $\ln(-\varepsilon)$. }
\end{figure}

In Fig.~\ref{fig:cfit2} we show $\heat(\varepsilon)$ and the curve
fitted to $-0.04<\varepsilon$ together with the relative error. The
fit is still not entirely convincing and the inset plot shows that
problems occur already at $\varepsilon<-0.01$. So, despite using three
terms the fit is still worse than the much simpler formula of our
first scenario in Fig.~\ref{fig:cfit1}.

\subsection{The low-temperature range}
\begin{figure}
  \includegraphics[width=0.483\textwidth]{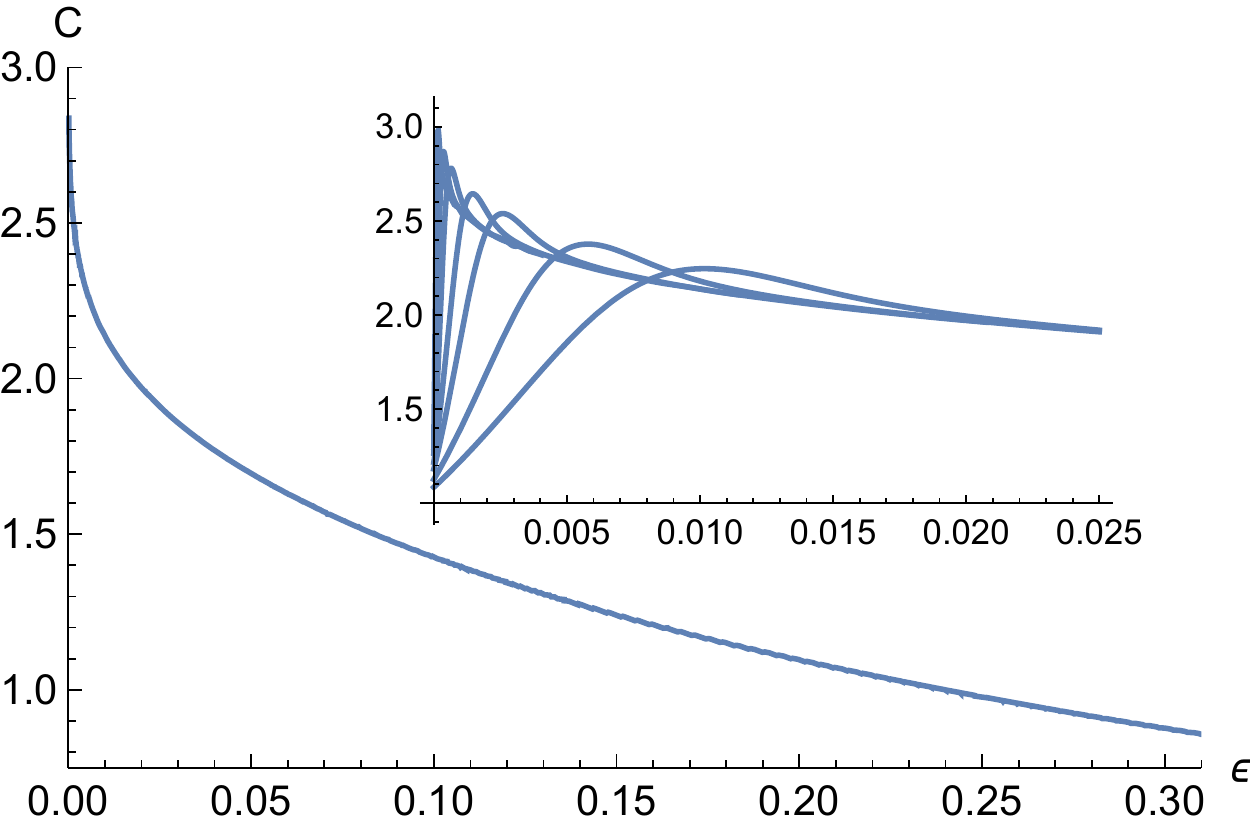}
  \caption{\label{fig:ceps2}(Colour on-line) Asymptotic specific heat
    $\heat(\varepsilon)$ versus $\varepsilon$ for low
    temperatures. Inset shows $\heat_L$ versus $\varepsilon$ for
    $L=6$, $8$, $12$, $16$, $24$, $32$, $48$, $64$, $96$. Asymptotic
    curve $\heat_{\infty}$ built from piecing together $\heat_L$ for
    $6\le L\le 160$.}
\end{figure}

Repeating this exercise on the low-temperature side ($\varepsilon>0$)
turns out to be somewhat more demanding. The data are more noisy and
the curves have a richer behaviour, making it harder to tell where the
asymptotic curve starts and ends for finite $L$. But, being a bit more
restrictive in our choices, we can piece together the curves for $6\le
L\le 160$ to obtain a rough limit curve for $0.149720\le K\le 0.2$,
or, $0.00018\lesssim\varepsilon\le 0.336$.  The resulting limit curve
and some finite examples are shown in Fig.~\ref{fig:ceps2}.

\begin{figure}
  \includegraphics[width=0.483\textwidth]{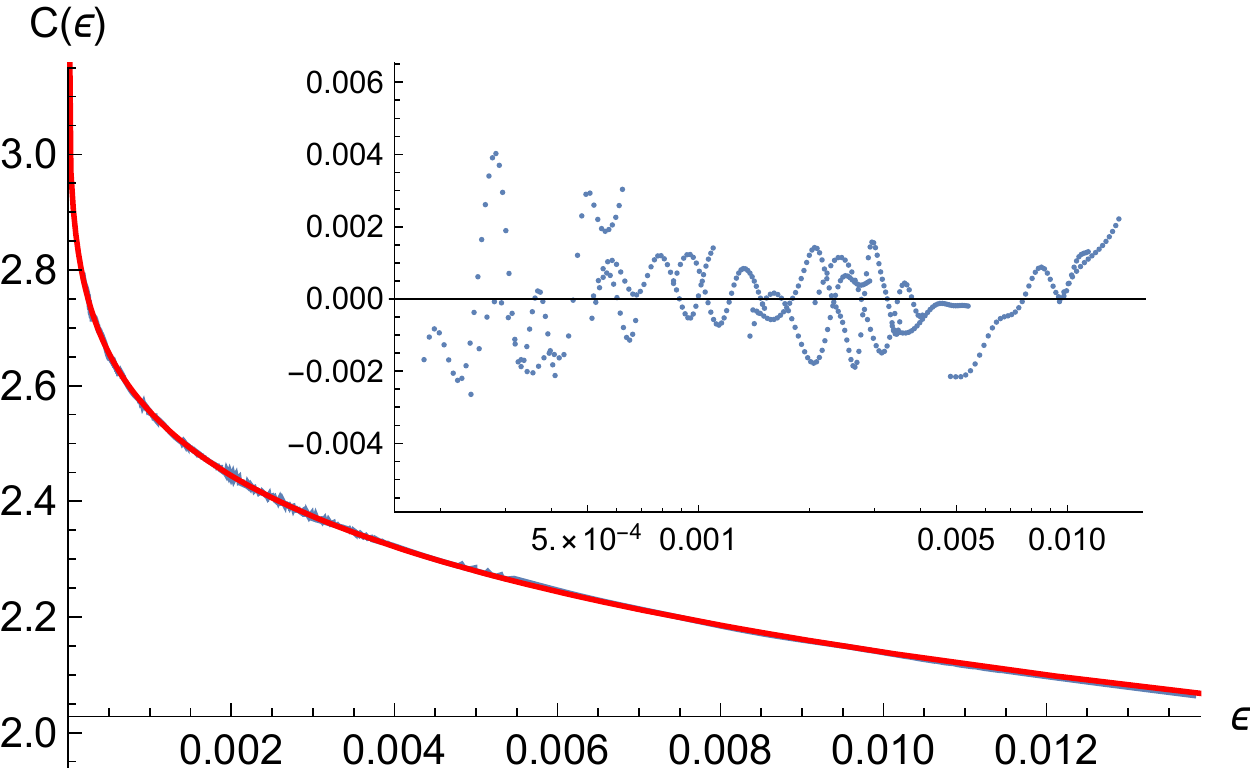}
  \caption{\label{fig:cfit3}(Colour on-line) Asymptotic specific heat
    $\heat(\varepsilon)$ and the fitted curve
    $y=3.60-2.856\varepsilon^{0.145}$ (indistinguishable) versus
    $\varepsilon$. Inset shows relative error $(y-\heat)/\heat$ versus
    $\ln(\varepsilon)$.}
\end{figure}

Fitting $y=A_0+A_1 \varepsilon^{\theta^+}$ over different intervals
works fine if we limit the data at an upper bound of
$\varepsilon<0.015$. Adjusting the upper bound between $0.010$ and
$0.015$ gives parameters $A_0=3.60(1)$, $A_1=-2.856(1)$ and
$\theta^+=0.145(2)$ and we show $y$ and $\heat$ in
Fig.~\ref{fig:cfit3}.  The relative error, shown in the inset, is
larger here than for $\varepsilon<0$ but not by much. 

\begin{figure}
  \includegraphics[width=0.483\textwidth]{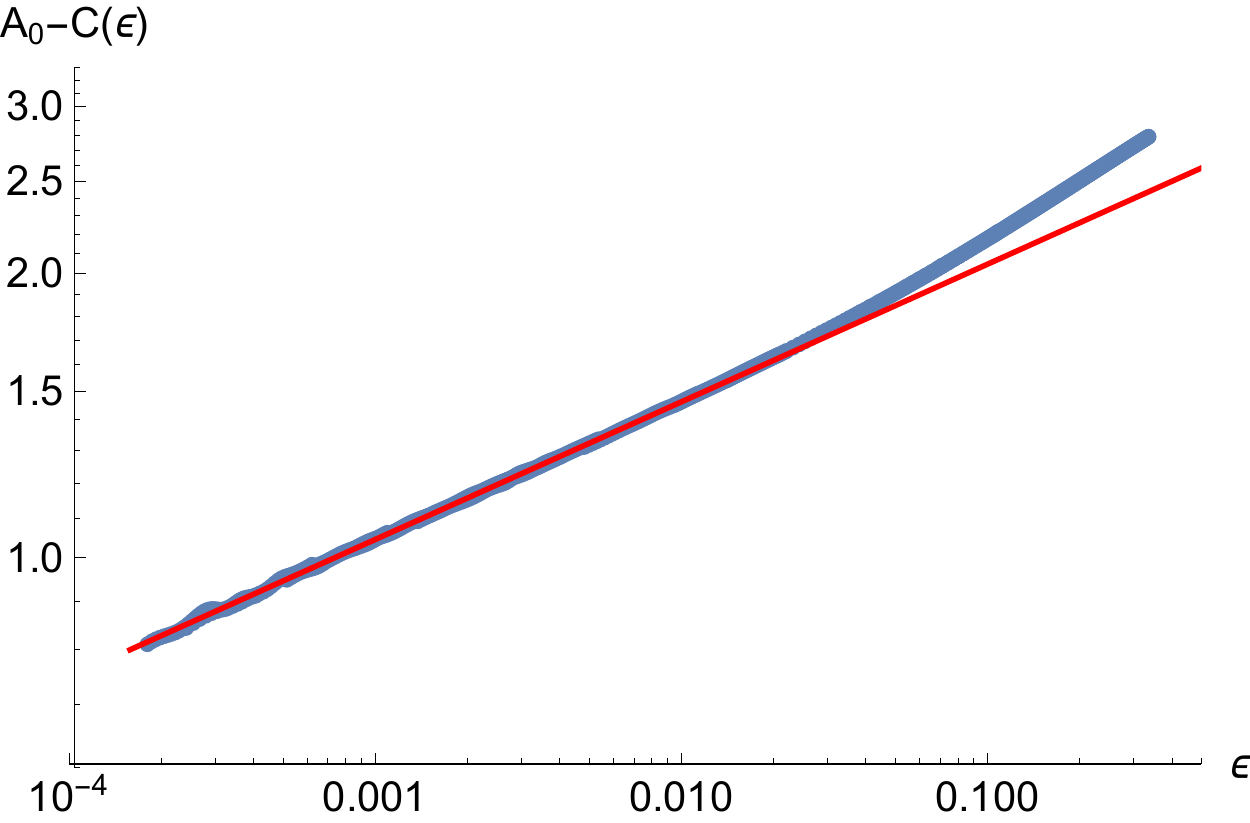}
  \caption{\label{fig:clog2}(Colour on-line) Log-log plot showing
    $3.60-\heat(\varepsilon)$ versus $\varepsilon$. The red line is
    $y=\ln(2.856)+0.145x$ where $x=\ln(\varepsilon)$.}
\end{figure}

The fitted curve then suggests an upper bound of $3.60$.  In
Fig.~\ref{fig:clog2} we show $3.60-\heat(\varepsilon))$ versus
$\varepsilon$ in a log-log plot, again demonstrating the linear
behaviour in such a plot.

\begin{figure}
  \includegraphics[width=0.483\textwidth]{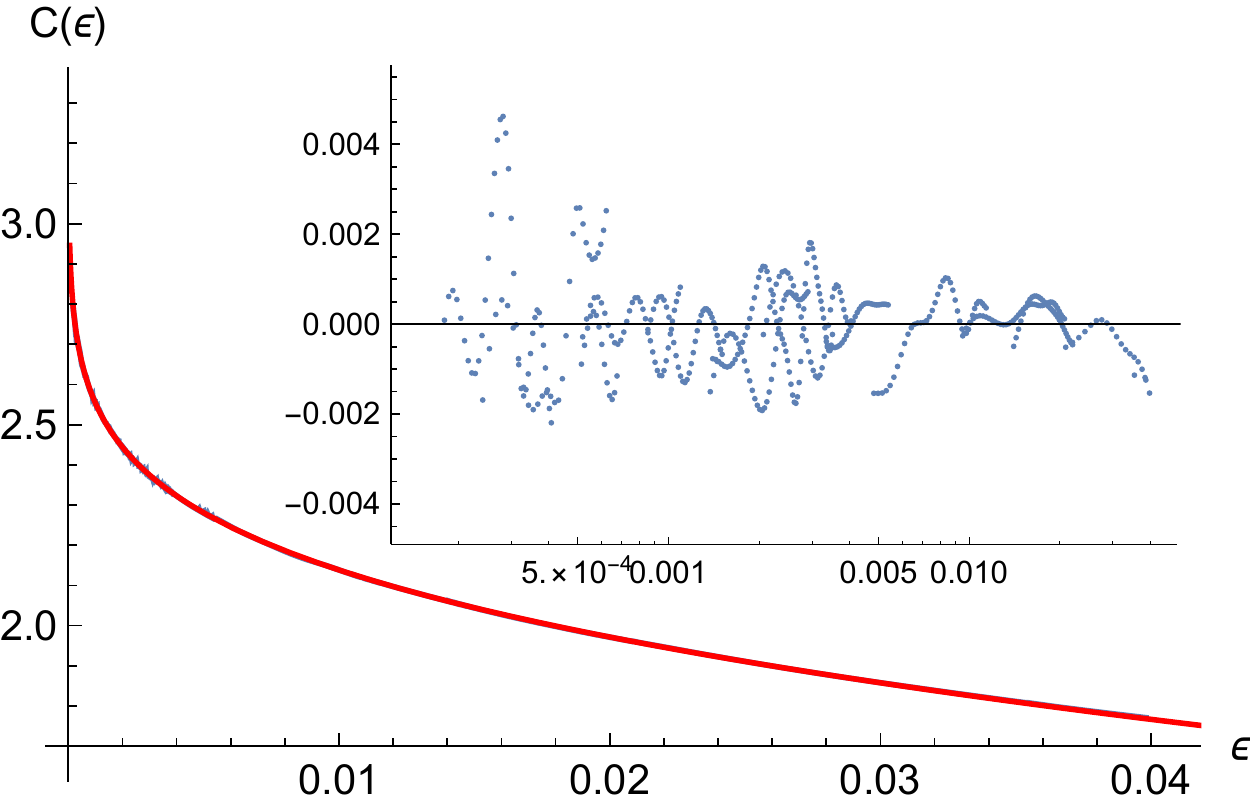}
  \caption{\label{fig:cfit4}(Colour on-line) Asymptotic specific heat
    $\heat(\varepsilon)$ and an indistinguishable fitted curve $y$ of
    Eq.~\eqref{yfit2} (see text for details) versus
    $\varepsilon$. Inset shows relative error $(y-\heat)/\heat$ versus
    $\ln(\varepsilon)$. }
\end{figure}

Next we try the divergent scenario on the low-temperature side as
well.  Fitting the three term expression of Eq.~\eqref{yfit2} works
quite well over a wide interval, whereas its two-term form fails to be
convincing for all intervals. Over $\varepsilon<0.04$ (changing the
upper bound to obtain error bars) we find $A_0=-1.85(2)$,
$A_1=1.851(3)$ and $A_2=1.61(3)$.  This is shown in
Fig.~\ref{fig:cfit4} and the fit indeed appears convincing as the
relative error shows.

As we have seen, in the canonical ensemble both scenarios can give as
plausible fits to the data.  However, both here and in the section on
finite-size scaling we consistently see cleaner fits for the
mean-field scenario, means less trends in the residual errors.  The
bounded scenario also gives a simpler model for the data, while the
divergent scenario requires three terms to work even with a modified
the higher-order correction term.

\subsection{A comparison with higher dimensions}
\begin{figure}
  \includegraphics[width=0.483\textwidth]{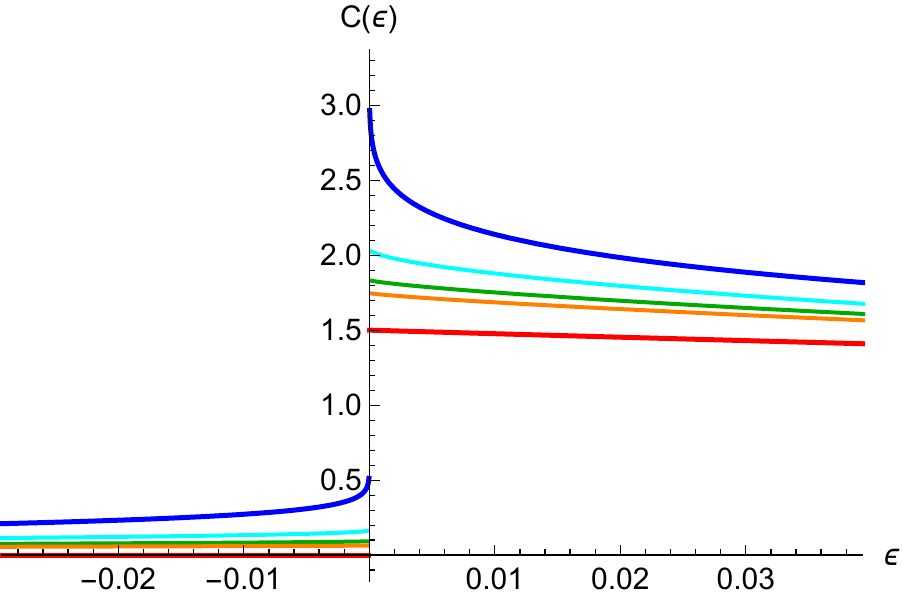}
  \caption{\label{fig:c45670}(Colour on-line) Asymptotic specific heat
    $\heat(\varepsilon)$ versus $\varepsilon$ for cases (downwards) 4D
    (blue), 5D (cyan), 6D (green), 7D (orange) and mean-field (red).}
\end{figure}

In Ref.~\cite{lundow:15} values for the two singular exponents
$\theta^-$ and $\theta^+$ which we have used for the bounded scenario,
as well as the limiting curves were found for larger dimensions. This
was done via Monte Carlo data for dimensions 5, 6 and 7, and was
derived exactly for the mean-field limit of the Ising model on
complete graphs.  Using data from that paper we show in
Fig.~\ref{fig:c45670} the suggested limit specific heat for systems
dimensions 4, 5, 6, 7 and the mean-field case. The exact form used for
$D=4$ is $\heat(\varepsilon)=3.60-2.86\varepsilon^{0.15}$ for
$\varepsilon>0$, $\heat(\varepsilon)=0.81-0.86(-\varepsilon)^{0.10}$
for $\varepsilon<0$ and $\heat(0)=1.84$.  As we can see, the behaviour
for $D=4$ is consistent with those for higher dimensions

\subsection{Inside the scaling window}

For $D\geq 5$ we know that the Ising model on the cyclic cubic lattice
Ref.~\cite{lundow:15a}, just like the model on complete graphs
Ref.~\cite{luczak:06}, have several nested scaling windows, inside of
which we see a non-trivial behaviour.  Assuming that the model for
$D=4$ follows the same pattern, having a bounded specific heat, we
here attempt to reconstruct the limiting specific heat for one of
these scaling windows, that given by taking $\kappa$ as the scaled
energy.

\begin{figure}
  \includegraphics[width=0.483\textwidth]{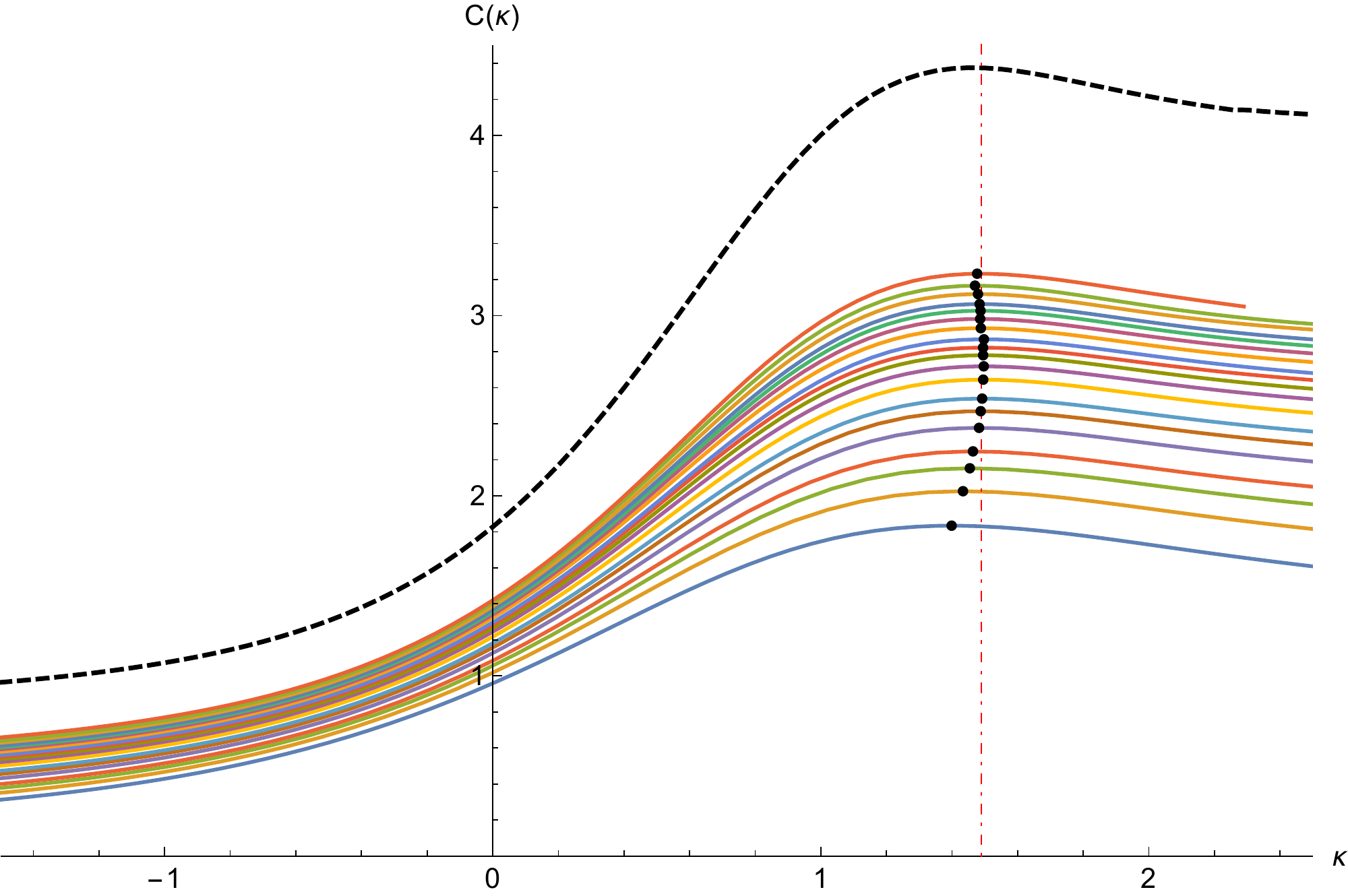}
  \caption{\label{fig:ckappa1}(Colour on-line) Limit specific heat
    $\heat(\kappa)$ (dashed) and $\heat_L(\kappa)$ for $L=6, 8,\ldots,
    256$ (upwards). Dots mark location of maximum for
    $\heat_L$. Dot-dashed vertical line is $\kappa=1.49$ of
    Fig.~\ref{fig:cmaxk}.}
\end{figure}

In Fig.~\ref{fig:ckappa1} we show the limit $\heat(\kappa)$ and the
finite-$L$ $\heat_L(\kappa)$ versus $\kappa=L^2(K-K_c)/K_c$.  Here we
use the estimate for the critical coupling $K_c=0.14969377$, but the
overall picture does not change much when changing $K_c$ in the 8th
decimal. The correction term from the maximum specific heat,
$0.224/L^2$, translates to $\kappa^*=0.224/K_c\approx 1.49$. The
asymptotic $\heat(\kappa)$ is obtained from fitting a line to
$\heat_L(\kappa)$ versus $1/L^{1/5}$ for $L\ge 32$ giving a rough
estimate of the limit, see Fig.~\ref{fig:ckc1}. The location of the
maximum of the limit $\heat(\kappa)$ curve almost matches the line
obtained from the finite-size maxima.  The maximum value of the limit
curve is $C_{\max}=4.37$, located at $K=1.49$, and the limit at
$\kappa=0$ is $\heat(0)=1.84$.

\section{Scaling for critical points and the value of $K_c$  }
In this section we will examine additional finite-size critical points
for the different energy moments and use them to better estimate the
critical temperature $K_c$

\subsection{The  specific heat}
\begin{figure}
  \includegraphics[width=0.483\textwidth]{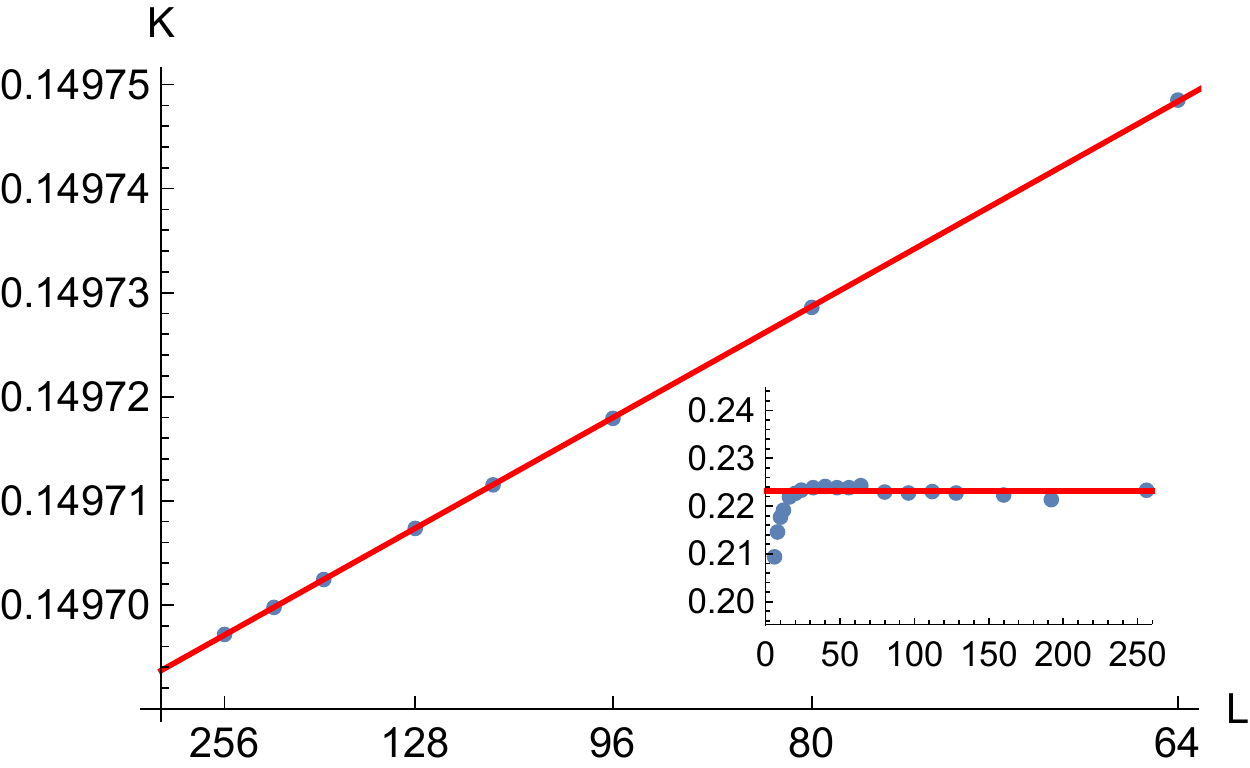}
  \caption{\label{fig:cmaxk}(Colour on-line) Location $K_c(L)$ of
    specific heat maximum versus $1/L^2$ for $L=64$, $80$, $96$,
    $112$, $128$, $160$, $192$, $256$.  The fitted line is
    $y=0.14969372(13)+0.224x$ where $x=1/L^2$. The inset shows
    $(K_c(L)-K_c)L^2$ versus $L$ and the constant line $y=0.224$.}
\end{figure}

In Fig.~\ref{fig:cmaxk} we show the location of the maximum specific
heat, $K_c(L)$, versus $1/L^2$ for $L\ge 64$. Fitting lines to
different $L$-ranges with $24\le L_{\min}\le 64$ gives
$K_c=0.14969372(13)$ where the error bar also matches the mean
difference between the line and the points for $L\ge 64$.  There are
problems with estimating this maximum. It is a wide and not very
distinct maximum and is thus sensitive to the slightest noise in the
data. It is also comparatively far from $K_c$ and the scaling is
rather slow (only $1/L^2$). Nevertheless, $1/L^2$ appears correct
since there is no systematic correction for large $L$, see inset plot
of Fig.~\ref{fig:cmaxk}.

It has been suggested that the location of such a pseudocritical point
should approach $K_c$ as $L^{-2}(\ln L)^{-1/6}$ \cite{lai:90,kenna:04}
but we find no such indication here. In fact, we let Mathematica fit
the points to an expression of the form $A_0+A_1L^{-2}+A_2L^{-2}(\ln
L)^{-1/6}$ but depending on which points are included in such a fit
the coefficient $A_2$ ends up being very close to zero.  Fitting for
example $L\ge 32$ and removing one point at a time to obtain an
average and an error bar we find $A_0=0.14969372(5)$, $A_1=0.218(7)$
and $A_2=0.004(5)$ (error bar from standard deviation), that is, $A_2$
is effectively zero with an error bar comparable to its mean
value. This also holds for other choices of $L$-ranges.

\begin{figure}
  \includegraphics[width=0.483\textwidth]{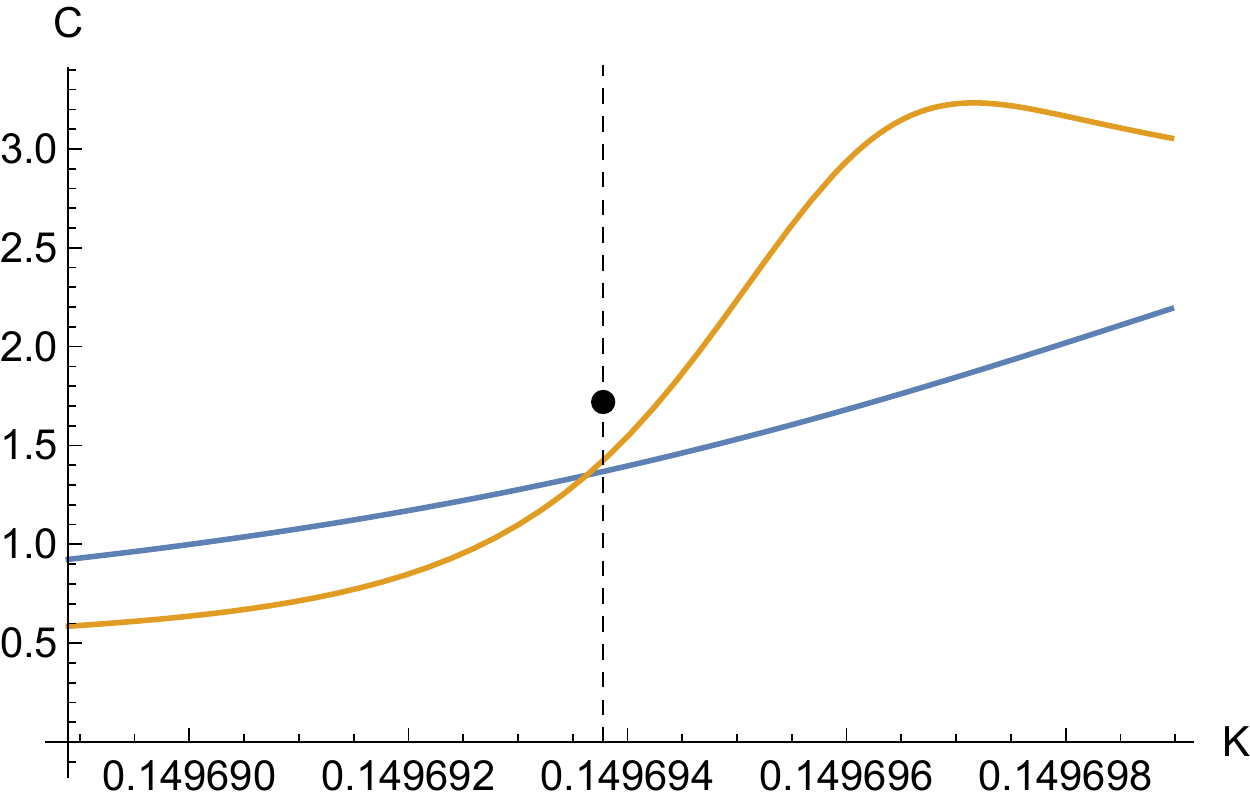}
  \caption{\label{fig:cx}(Colour on-line) Specific heat $\heat_L(K)$
    and $\heat_{2L}(K)$ (steeper) for $L=128$ indicating a distinct
    $(128,256)$ crossing point $K_x(128,256)=0.14969363$ close to
    $K_c$ (dashed line). The point is at an estimated limit of
    $\heat_L(K_c)$ when $L\to\infty$.}
\end{figure}

The finite-size behaviour of the specific heat also provides a third,
and more powerful, technique for estimating $K_c$, namely to use
$(L,2L)$-crossing points of the specific heat.  Let $K_x=K_x(L_1,L_2)$
denote the point where $\heat_{L_1}(K_x)=\heat_{L_2}(K_x)$. In general
one could use $(L,bL)$ crossing points but since our data give the
largest number of such pairs for $b=2$ we will only use these. In
Fig.~\ref{fig:cx} we show an example of such a crossing point for the
case $(128,256)$. An advantage of these points is that they are quite
distinct and occur very close to $K_c$.

\begin{figure}
  \includegraphics[width=0.483\textwidth]{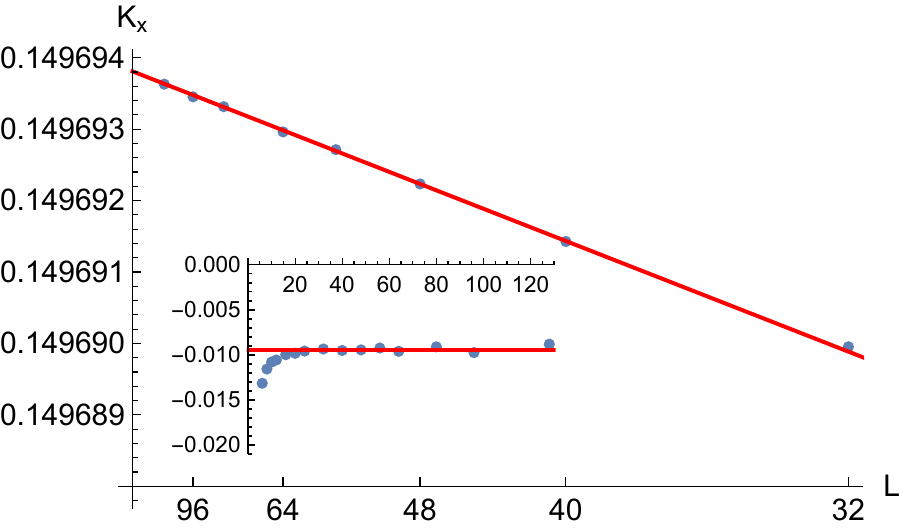}
  \caption{\label{fig:ckx}(Colour on-line) Specific heat crossing
    points $K_x(L,2L)$ versus $1/L^{9/4}$ for $L=32$, $40$, $48$,
    $56$, $64$, $80$, $96$, $112$ and $128$. The fitted line is
    $0.14969379-0.0093x$ where $x=1/L^{9/4}$. The inset shows the
    $(K_x-0.14969379)L^{9/4}$ and the line $y=0.0093$.}
\end{figure}

In Fig.~\ref{fig:ckx} we show $K_x(L,2L)$ for $32\le L\le 128$ and a
fitted line giving $K_x=0.14969379(1) - 0.0093(1)\,L^{-9/4}$.  The
error bars are, as before, obtained from fitting lines to $K_x$ versus
$1/L^{9/4}$ for $L$-ranges with $24\le L_{\min}\le 64$ and taking the
mean difference between a fitted line and the points for $L\ge 32$.
In fact, this technique appears quite robust and does not depend
strongly on which points are included in the fits. A huge benefit is
that the correction term seems to be of order $1/L^{9/4}$ with
negligible higher-order correction terms. Unfortunately we do not have
a theoretical basis for this exponent. We think it is supported by
looking at the correction $(K_x-K_c)L^{9/4}$ which is almost constant
$-0.0093$ for $L\ge 16$, see inset of
Fig.~\ref{fig:ckx}. Alternatively we can express the rescaled crossing
point as $\kappa_x\approx -0.062/L^{1/4}$ to clearer show that this
point goes to zero also in the scaling window.

The value of $\heat_L(K_x)$ scales very much like that of
$\heat_L(K_c)$ of Fig.~\ref{fig:ckc1}. We estimate $\heat_L(K_x) =
1.838(2)-1.29(1)/L^{1/5}$ thus giving a slightly different correction
term but approximately the same limit (not shown) though there is more
noise here than for $\heat_L(K_c)$.

\subsection{Energy kurtosis}

\begin{figure}
  \includegraphics[width=0.483\textwidth]{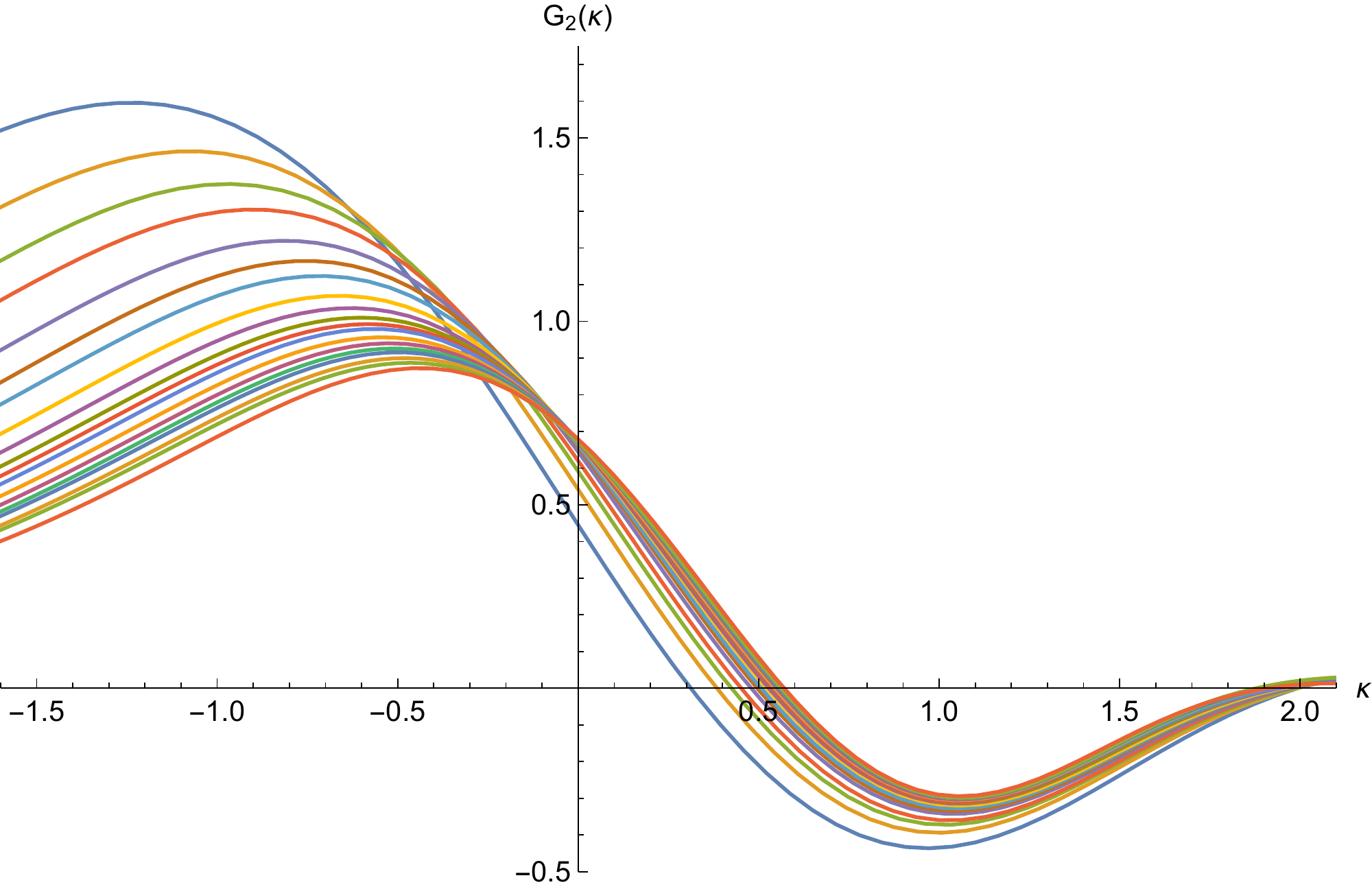}
  \caption{\label{fig:g2kappa}(Colour on-line) Excess kurtosis
    $\ekurt(\kappa)$ for $6\le L\le 256$.}
\end{figure}

In Fig.~\ref{fig:g2kappa} we show the energy (excess) kurtosis
$\ekurt_L(\kappa)$ for all $L$. There is a distinct minimum near
$\kappa=1$ and a maximum approaching from the left.  Shortly we will
estimate the scalings for these critical points.

\begin{figure}
  \includegraphics[width=0.483\textwidth]{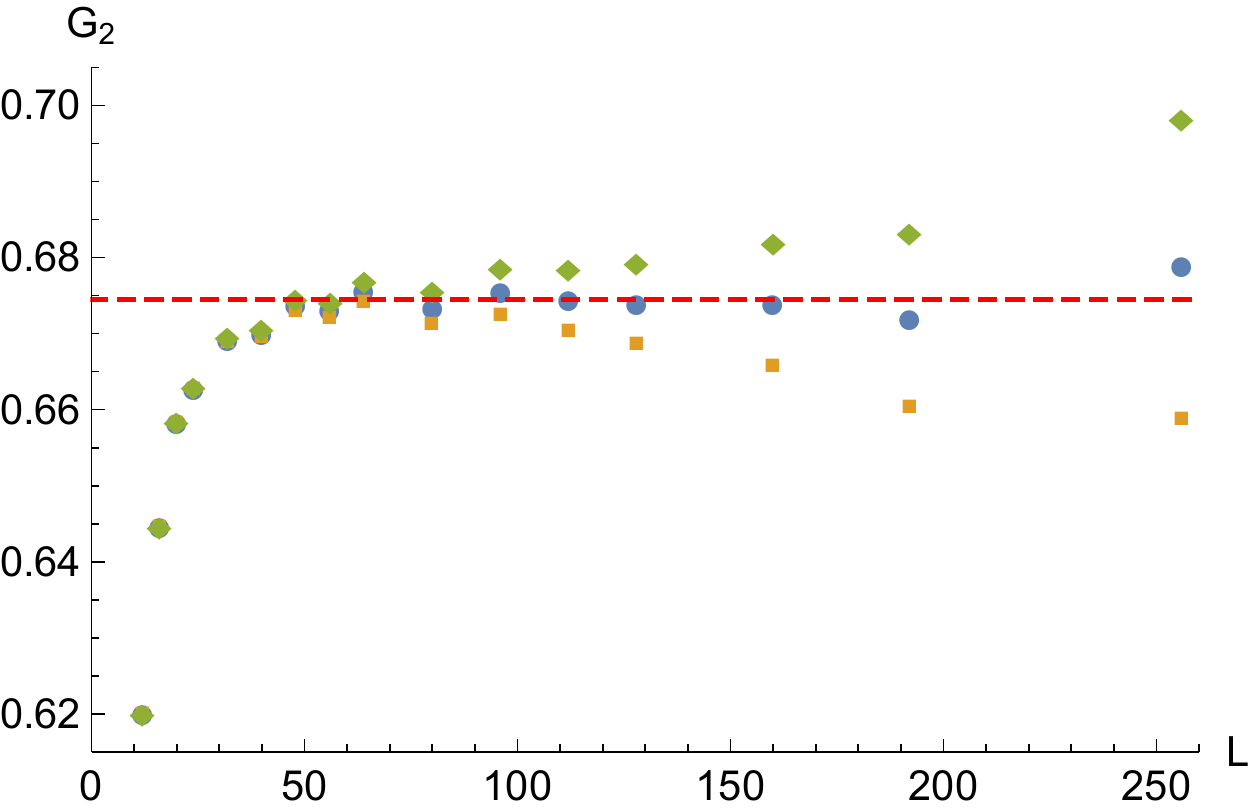}
  \caption{\label{fig:g2kc}(Colour on-line) Kurtosis $\ekurt_L(K_c$
    versus $L$ for $10\le L\le 256$ with $K_c=0.149693785$. The dashed
    constant line through the points is $y=0.674$ (fitted on $L\ge
    48$). The points trending downwards and upwards are for $K_c\pm
    5\times 10^{-8}$.  }
\end{figure}

We will, however, start with the value at $K_c$ as it turns out to be
an excellent arbiter of $K_c$. In Fig.~\ref{fig:g2kc} we show
$\ekurt_L(K_c)$ versus $L$ with $K_c=0.149693785$ and also for $K_c\pm
5\times 10^{-8}$ where the points clearly trend up or down. In fact, 
this effect is distinctly visible already for $\pm 2\times 10^{-8}$
which means that we can set $K_c=0.149693785(20)$. This is in perfect
agreement with the result in relation to Fig.~\ref{fig:ckc2}. Obviously
some noise sets in for $L=192, 256$ but we have based the present
$K_c$ on $L\le 160$ giving very small fluctuations ($\pm 0.001$)
around the line.

Let us now continue with the maxima and minima of
Fig.~\ref{fig:g2kappa}. Let first $K_c(L)$ be the local minimum of
$\ekurt_L$ located near $\kappa=1$.  This minimum is quite distinct
and we thus expect a good estimate of $K_c$. Indeed, fitting as before
$K_c(L)=A_0+A_1/L^2$ to data ranges $L_{\min}\le L\le 256$ with $32\le
L_{\min}\le 128$ we find $K_c=A_0=0.14969380(2)$ and $A_1=0.1579(4)$
(from median and interquartile range), a surprisingly sharp estimate
of $K_c$ from this kind of critical point. Fitting instead
$y=A_0+A_1/L^a$ does not change the result much. We then obtain
$K_c=A_0=0.14969378(4)$, $A_1=0.156(9)$ and $a=2.00(1)$, quite
consistent with the previous only with larger error bars. Note in
particular the $a$-estimate, showing no sign of any logarithmic
correction.

\begin{figure}
  \includegraphics[width=0.483\textwidth]{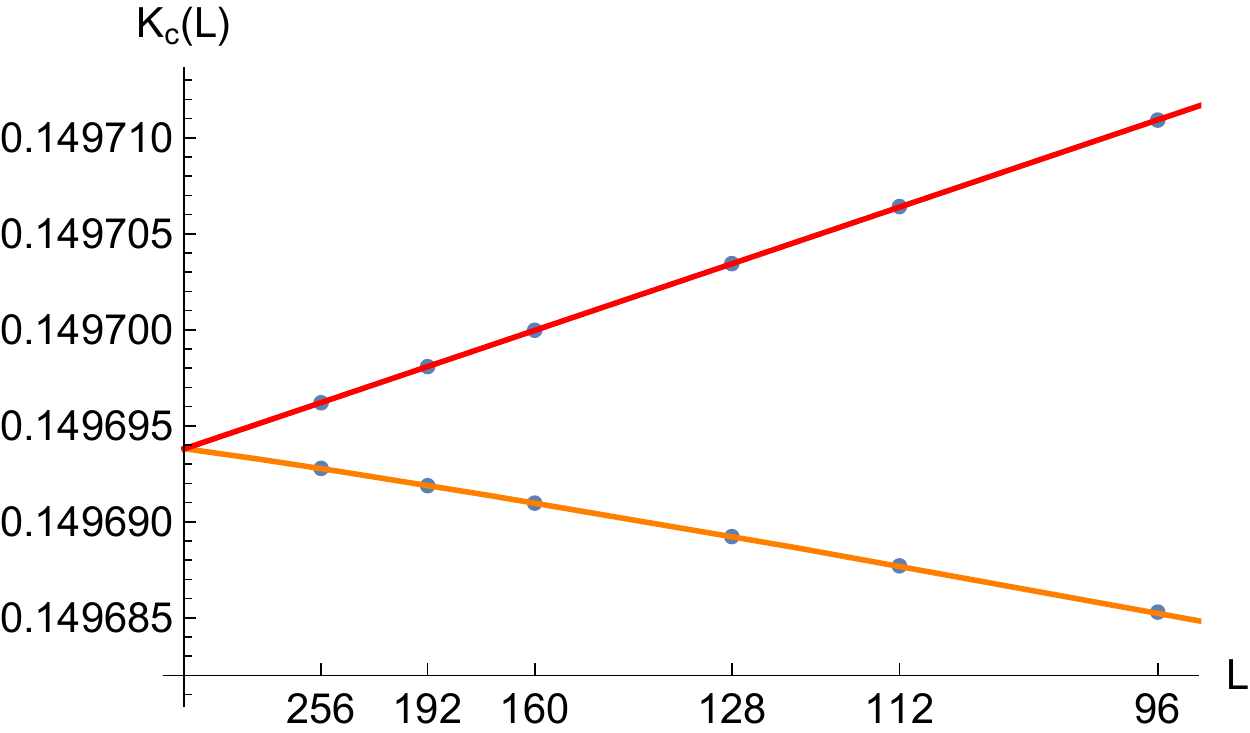}
  \caption{\label{fig:g2maxmink}(Colour on-line) Location $K_c(L)$ of
    the maximum (from below) and minimum (from above) kurtosis versus
    $1/L^2$ for $L=96$, $112$, $128$, $160$, $192$, $256$. The red
    line is $y=0.14969380(2)+0.1579(3)x$ and the orange curve is
    $y=0.14969381(5)-0.051(5)x-0.27(5)x^{5/4}$, where $x=1/L^2$. See
    text for information on error bars.}
\end{figure}

For the location $K_c(L)$ of the maximum we need a correction
term. Fitting $K_c(L)=A_0+A_1/L^2+A_2/L^a$ to data ranges $20\le
L_{\min}\le 64$ gives fairly stable results: $A_0=0.14969380(5)$,
$A_1=-0.054(13)$, $A_2=-0.31(20)$ and $a=2.56(26)$. Fixing instead
$a=2.5$ and repeating for $32\le L_{\min}\le 128$ we obtain
$K_c=A_0=0.14969381(5)$, $A_1=-0.051(5)$ and $A_2=-0.27(5)$,
consistent with the previous free-$a$ estimate. One might here
consider using a correction term of the form $A_2L^{-2}(\ln L)^{-1/6}$
but the result then depends strongly on $L_{\min}$ with clear trends
in the $A_i$. Using instead a free exponent $a$ of $\ln L$ gives too
much noise in $a$ to say what its value is. This suggests that the
data do not favor a correction on this form. In
Fig.~\ref{fig:g2maxmink} we show $K_c(L)$ for both the maximum and the
minimum and the fitted curves. In rescaled form they can now be
expressed as $\kappa^*=1.055(3)$ and $\kappa^*=-0.34(3) -
1.8(3)/\sqrt{L}$ for the minimum and maximum respectively.

We shall also attempt to estimate the scaling of the maximum and
minimum values.  First we need to find the correct leading order
correction. Fitting $\ekurt=A_0+A_1/L^a$ to $32\le L_{\min}\le
64$ for the maxima gives $a=0.50(4)$ but for the minima we
obtain $a=0.60(15)$. There is, however, more support for a
leading order correction term $A_1/\sqrt{L}$.  

Having selected $a=1/2$ we fit the two-term formula to the minima for
$32\le L_{\min}\le 96$ and obtain
$\ekurt_{\min}=-0.2817(5)-0.209(6)/\sqrt{L}$. For $L<32$ some
correction term becomes necessary though but we do not know what form
it should take. The same procedure applied to the maxima gives us
$\ekurt_{\max}=0.764(1)+1.72(1)/\sqrt{L}$, again with higher-order
corrections needed for $L<32$.

\begin{figure}
  \includegraphics[width=0.483\textwidth]{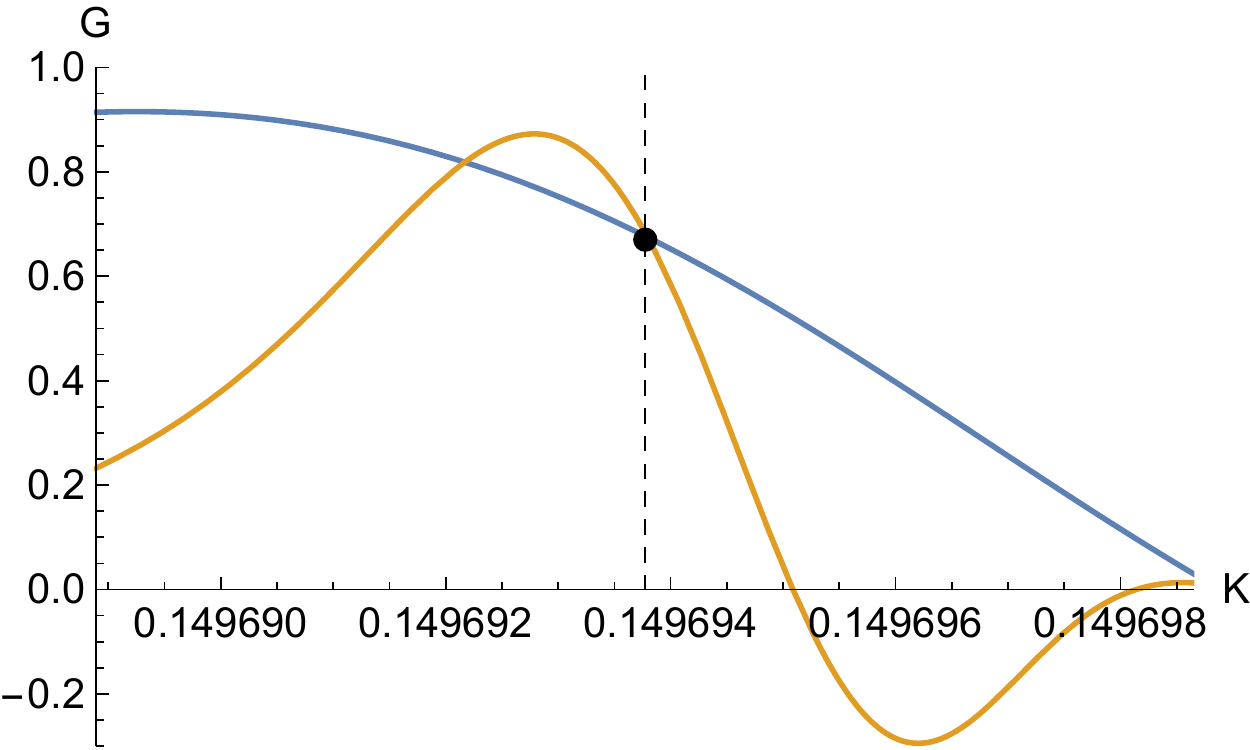}
  \caption{\label{fig:g2x}(Colour on-line) Kurtosis $\ekurt_L$ and
    $\ekurt_{2L}$ for $L=128$ indicating a distinct $(128,256)$
    crossing point close to $K_c$ (dashed line). Here
    $K_x(128,256)=0.149693796$. The point is at the estimated limit of
    $\ekurt_L(K_c)=0.675$ when $L\to\infty$.  }
\end{figure}

\begin{figure}
  \includegraphics[width=0.483\textwidth]{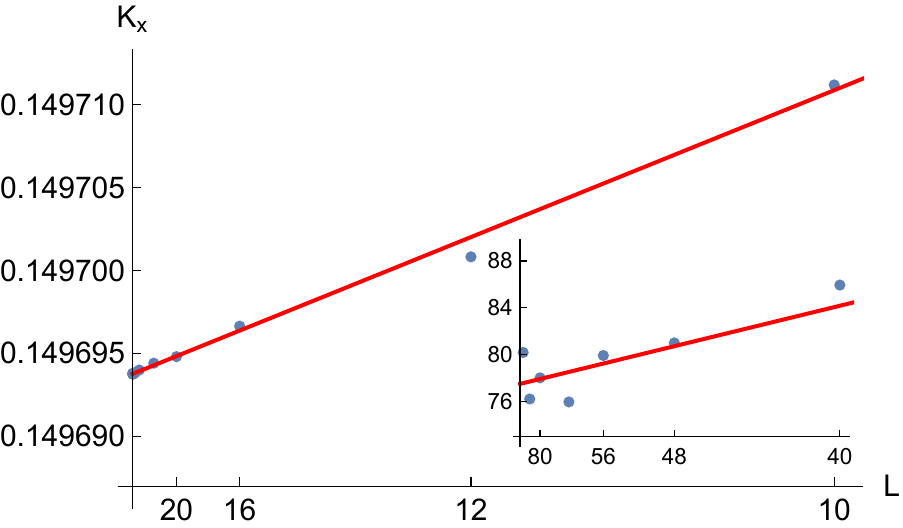}
  \caption{\label{fig:g2kx}(Colour on-line) Kurtosis crossing point
    $K_x(L,2L)$ versus $1/L^4$ for $10\le L\le 128$. The red line is
    $y=0.149693775(15) + 0.17(4)x$ where $x=1/L^4$. Inset shows $K_x$
    for $L=40$, $48$, $56$, $64$, $80$, $96$, $128$ and the same line.
    The numbers on the $y$-axis are the decimals after
    $0.149693\ldots$.}
\end{figure}

We shall end the subject of kurtosis by studying its $(L,2L)$-crossing
point, denoted $K_x$. In Fig.~\ref{fig:g2x} we show an example for
$L=128$. What is striking with this crossing point is how extremely
close it is to $K_c$. We find the very small difference
$K_x-K_c\propto 1/L^4$. In Fig.~\ref{fig:g2kx} we show the crossing
point $K_x$ versus $1/L^4$ for $L\ge 10$ and an inset for the larger
$L$. Fitting $K_x=A_0+A_1/L^4$ for $10\le L_{\min}\le 56$ gives
$K_c=A_0=0.149693775(15)$ and $A_1=0.17(4)$. The points appear more
scattered for $L\ge 64$ but they stay within $\pm 2\times 10^{-8}$
from the line.

\subsection{Energy}

We have obtained many estimates of $K_c$ of varying precision
depending on the studied quantity and approach. The crossing point of
the specific heat gave a sharp estimate of $K_c$. The energy kurtosis
turned out to be extremely useful for pinning down $K_c$ from several
different perspectives; local maxima and minima, crossing points,
constant kurtosis for large $L$.  On the other hand, the local maximum
of the specific heat was less useful, giving comparatively wide error
bars. Taking an intersection of all these estimates and using the
smallest of the error bars we find $K_c=0.149693785(10)$. This also
agrees with the recent~\cite{deng:2020} $K_c=0.14969388(22)$.  Our old
estimate~\cite{lundow:09} of $K_c=0.1496947(5)$ was then off by less
than two of its error bars.

\begin{figure}
  \includegraphics[width=0.483\textwidth]{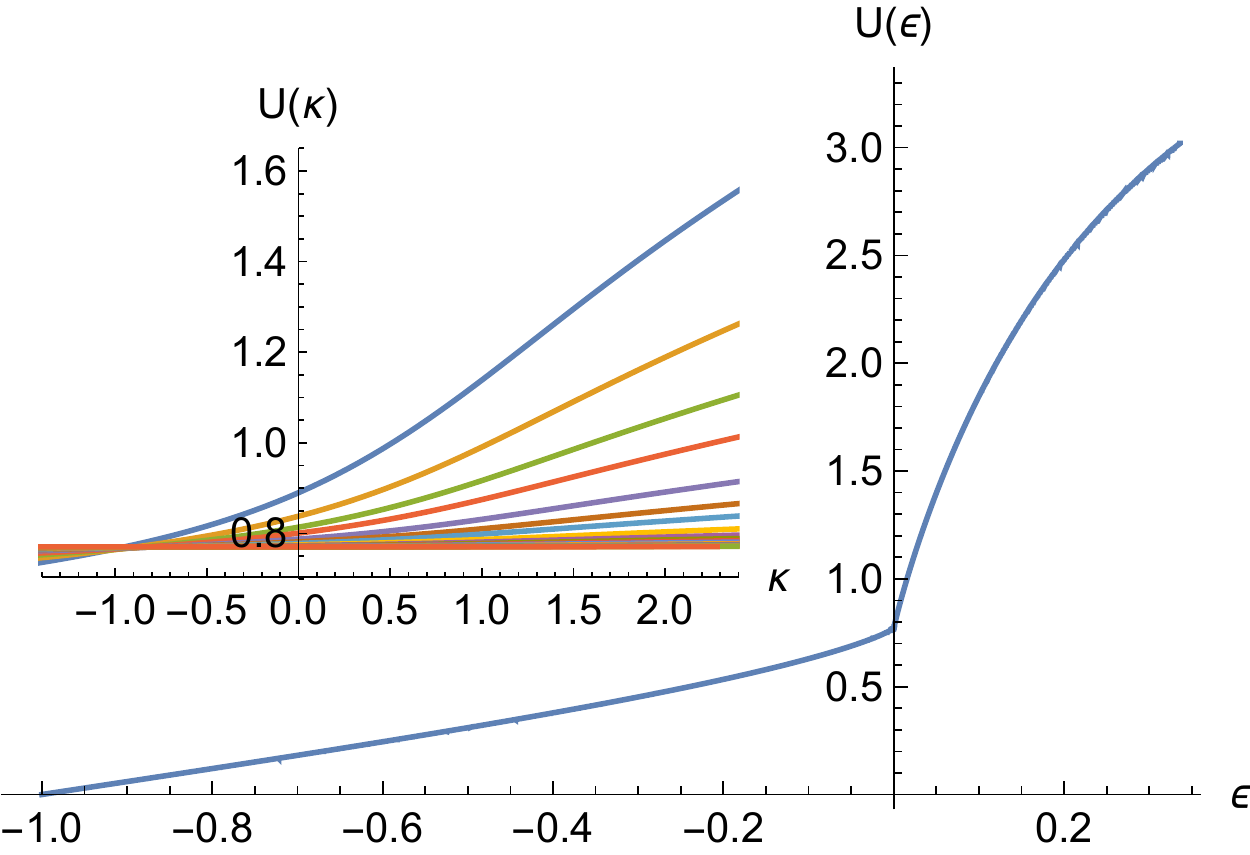}
  \caption{\label{fig:ueps}(Colour on-line) Asymptotic energy
    $\erg(\varepsilon)$ versus $\varepsilon$. Inset shows $\erg$
    versus reduced coupling $\kappa$ for $6\le L\le 256$.}
\end{figure}

The asymptotic energy $\erg(\varepsilon)$, obtained exactly as we did
above for the specific heat, is shown in Fig.~\ref{fig:ueps}. The
inset shows the energy versus the rescaled coupling for a range of
system sizes.  Clearly their limit approaches some constant inside the
scaling window. Fitting $y=A_0+A_1/L^2$ to a sequence of
$\erg_L(\kappa)$ for a fixed $\kappa$ is of course one
possibility. Unfortunately this requires higher order corrections and
it is not clear what exponent a further term should use. Including a
third term $A_2/L^a$ gives very unstable results at $\kappa=0$.

\begin{figure}
  \includegraphics[width=0.483\textwidth]{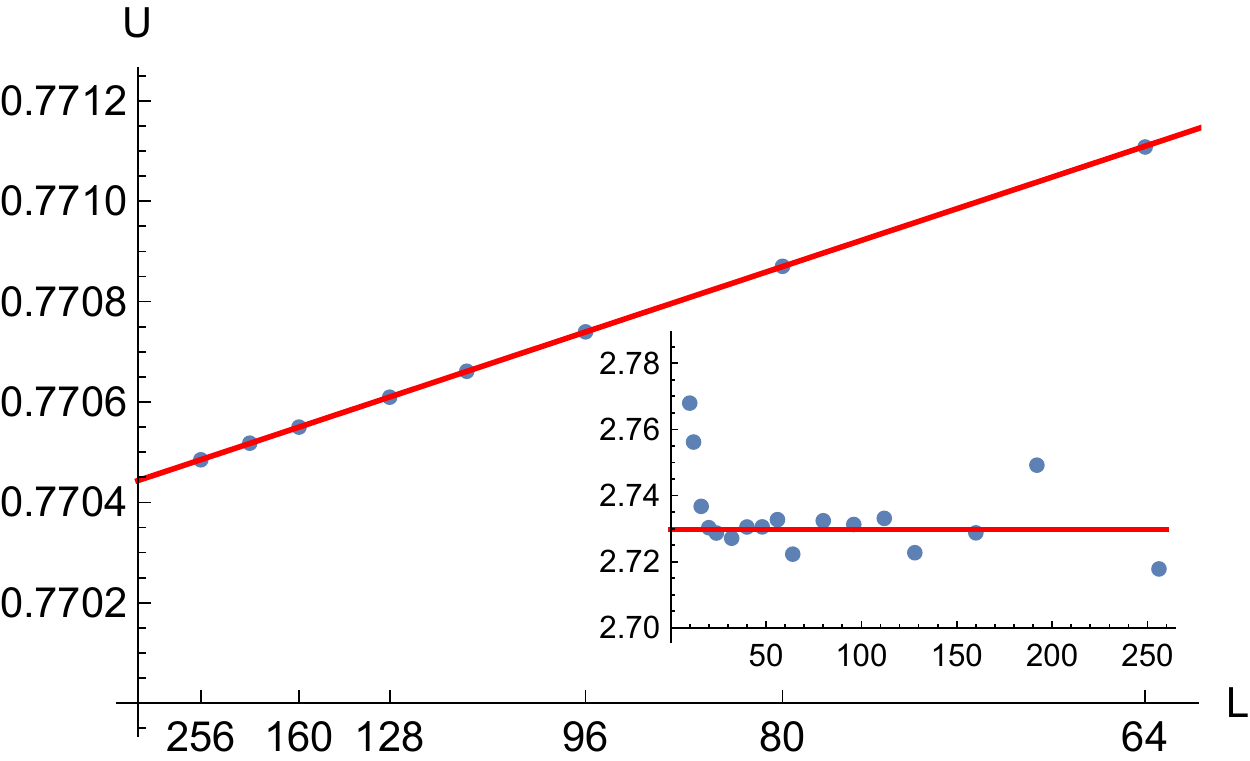}
  \caption{\label{fig:ukappa}(Colour on-line) Energy $\erg_L(\kappa)$,
    where $\kappa=-0.25$, versus $1/L^2$ for $64\le L\le 256$.  The
    red line is $y=0.7704434(10)+2.730(2)x$ (see text) with $x=1/L^2$.
    Inset shows $(\erg_L(\kappa)-0.7704434)L^2$ versus $L$ and the
    line $y=2.73$.}
\end{figure}

However, if we let $\kappa=-0.25$ then the higher-order corrections
effectively vanish. In Fig.~\ref{fig:ukappa} we show
$\erg_L(\kappa=-0.25)$ versus $1/L^2$ and the inset shows only
noise-like corrections. The result is $\erg_c=A_0=0.7704434(10)$ and
$A_1=2.730(2)$ where error bars indicate mean difference between line
and points and standard deviation in fits over $20\le L_{\min}\le 96$,
respectively.

\section{The Universality Class}
After a full analysis of our data from both the microcanonical and
canonical point of view we are now ready for the main question of this
paper: Which universality class does the 4-dimensional Ising model
belong to?  The existing literature has actually considered three
possibilities: mean-field behaviour with bounded specific heat, a
logarithmic singularity in the specific heat of the type as in the
$\phi^4$-model, or a weak first order phase transition.

As mentioned in the introduction, assuming that we have a
$\phi^4$-singularity one is led to expect a very slow divergence, see
Eq. \eqref{crule1} of the specific heat as a function of $L$, the term
$(\ln L)^{1/3}$ merely increases from 1.4 to 1.9 when $L$ increases
from 16 to 1024.  Because of this one would not expect to clearly
distinguish this case from one with bounded heat via classical Monte
Carlo simulation.  As we have seen in Sections \ref{finscan} and
\ref{2scen} the data for the canonical ensemble can be fitted to the
finite-size scaling and the asymptotic form expected from the
$\phi^4$-case. However, for both scaling and asymptotics we get
smaller errors when using the scaling and asymptotic forms given by
the scenario with a bounded specific heat.  So, as expected the
difference is not clear, but a bounded specific heat consistently
fares better.

However, when instead examining the model in the microcanonical
ensemble the difference is clear cut.  In Section \ref{sec:micro} we
found that for the full range of $L$ used here the minimum value of
$\fK'(U)$ is positive and increasing to a finite, positive limit as a
function of $L$.  The only assumption underlying this conclusion is
that the Ising model in dimension 4 does not have finite-size effects
so large that lattices with $L\leq 256$ do not even present the right
direction of change for $\fK'(U)$ as function of $L$, something which
would be so far unheard of.

First, the fact that $\fK'$ is positive means that we have a positive
specific heat.  This in turn means that, unlike the situation in
dimension 5 \cite{lundow:11}, we do not have a meta-stable region for
finite $L$ and so a purely second order phase transition.  Here we can
conclude that the weak first-order phase transition suggested in
\cite{PhysRevD.100.054510,Akiyama_2020} is ruled out, even in the
quasi-first order form found in \cite{lundow:11}. In the thermodynamic
limit a first-order phase transition is already ruled out by the
rigorous results in \cite{aizenman,MR821310}.  The indications for a
first-order phase transition came from a numerical renormalisation
group technique which involves both a finite lattice size $L$ and a
cut-off parameter $D$. So, even if the method is sound this could
indicate that the accuracy as a function of these two parameters is
less than expected. This would also agree with the fact that the
estimate for $K_c$ in \cite{PhysRevD.100.054510} deviates noticeably
from the currently best.  The suggested first-order transition has
been used in some works on the Higgs field \cite{CC2020, Consoli_2021}
which might instead be redone with a mean-field transition in mind.

Second, and more importantly the fact that the asymptotic value of
$\fK'(U)$ is strictly positive means that we have a finite upper bound
for the specific heat in the thermodynamic limit.  This rules out a
singularity of the same type as in the $\phi^4$-model, and of course
any other form of divergent specific heat.  In Section \ref{2scen} we
saw that, just as for higher dimension, we could fit the canonical
data well to a discontinuous specific heat curve with distinct left
and right-side limits at $K_c$.  The values we obtained there also
agreed well with the bounds found in Section~\ref{sec:micro} for those
limits, via the estimated values for $\fK'(U)$. Here we can also note
that the limit values for the specific heat implies that the high- and
low-$U$ limits for $\fK'(U)$ should be 0.0277 and 0.00622, which is
compatible with the data in Fig.~\ref{fig:kdlim}.  If the model
follows the behaviour seen in higher dimension~\cite{lundow:11} then
one expects $\fK'(U)$ to approach these limit values controlled by a
pair of high- and low-$U$ singular exponents, and we have found good
fits of that form to our data.

Together this leads to the conclusion that the Ising model and the
$\phi^4$-model do not belong to the same universality class.  Instead,
the first has a mean-field type bounded discontinuous jump in the
specific heat at the critical point, and the second a logarithmically
divergent specific heat at the same point. 

Following the sometimes used terminology that the logarithmic singularity of the $\phi^4$-model is a "correction"  
to  mean field behaviour one could still claim that the two models belong to the same universality 
class, albeit with and without corrections. However, in our view this is not consistent with the generally 
accepted idea that universality classes correspond to fixed points of the renormalization group flow.  Since renormalisation 
does detect the logarithmic singularity of the 2D Ising model the method should be sensitive to such 
singularities more generally, as long as one works with a large enough subspace for the flow. Hence one should expect 
separate fixed points for the $\phi^4$-model and a pure mean-field singularity.   One might also simply think 
 that calling the leading order part of the singularity in 4-dimensions a "correction"  appears inaccurate.

\section{Discussion}
Our main conclusion here, that the Ising model and the $\phi^4$-model
belong to distinct universality classes for $D=4$ will be surprising
to some, but at the same time does not invalidate any major
theoretical tools or methods from the physics literature.  Rather, it
enriches the picture since we now have two models which at their upper
critical dimension have the same, mean-field, critical exponents but
still have distinct lower order behaviour.

As mentioned in the introduction, having the same spatial dimension
and symmetries is merely the simplest candidate for a list of
properties defining a universality class.  One of the major heuristic
underpinnings for the existence of universality classes is the
existence of fixed points for the renormalisation group flow on the
space of models, or hamiltonians, and here a split into two
universality classes would simply mean that there is some new variable
which differentiates two distinct fixed points for this flow.  Today
we know \cite{Enter:93} the renormalisation is a far more sensitive,
and often precarious, procedure than the early works assumed and it
would not be surprising if new features are relevant exactly at the
upper critical dimension.

At the moment we cannot answer what this new differentiating variable
corresponds to, but at a first glance there are two features which
distinguish the Ising and $\phi^4$-model from each other.  First, the
cardinality of the single-spin state space differentiates the Ising
model from the $\phi^4$-model for all finite parameters
$(b,\lambda)$. Second, is the fact that the spin values in the
$\phi^4$-model are unbounded, something which we alluded to already in
the introduction in connection with how the $\phi^4$-model may be
approximated by block-spins based on the Ising model.  The
$\phi^4$-model with spin values restricted to a fixed finite interval
$(-a,a)$ might provide an interesting intermediate case which could
rule out the first of these properties as the relevant one.

In this paper we have focused on energy related properties of the
model, with emphasis on the specific heat, but in our sampling runs we
have also collected data for the magnetisation. This has been done so
that both the canonical and microcanonical magnetisation distributions
can be reconstructed and we will provide an analysis for both $D=4$
and $D=5$ in an upcoming paper.  That analysis will shed  more light on 
the scaling limit of the Ising model and the related quantum field theory.

\begin{acknowledgments}
  The computations were performed on resources provided by the Swedish
  National Infrastructure for Computing (SNIC) at High Performance
  Computing Center North (HPC2N) and at Chalmers Centre for
  Computational Science and Engineering (C3SE).
\end{acknowledgments}


\end{document}